\documentclass{emulateapj}

\slugcomment{Accepted to The Astrophysical Journal}

\shorttitle{The Early Chemical Evolution of the Scuptor Group Dwarfs}
\shortauthors{Yang, S.-C. et al.}

\begin{document}


\title{The Early Chemical Enrichment Histories of Two Sculptor Group Dwarf Galaxies as Revealed by RR Lyrae Variables}


\author{Soung-Chul Yang\altaffilmark{1, 2, +}, Rachel Wagner-Kaiser\altaffilmark{3}, Ata Sarajedini\altaffilmark{3}, Sang Chul Kim\altaffilmark{1}, and  Jaemann Kyeong\altaffilmark{1}}

\affil{\altaffilmark{1}Korea Astronomy and Space Science Institute (KASI), Daejeon, 305-348, South Korea}
\affil{\altaffilmark{2}The Observatories of the Carnegie Institution for Science, 813 Santa Barbara Street, Pasadena, CA 91101, U
SA}
\affil{\altaffilmark{3}Department of Astronomy, University of Florida, P.O. Box 112055, Gainesville, FL 32611}

\email{sczoo@kasi.re.kr}


\altaffiltext{+}{KASI-Carnegie Fellow}


\begin{abstract}
We present the results of our analysis of the RR Lyrae (RRL) variable stars detected in two transition-type dwarf galaxies (dTrans), ESO294-G010 and ESO410-G005 in the Sculptor group, which is known to be one of the closest neighboring galaxy groups to our Local Group. Using deep archival images from the Advanced Camera for Surveys (ACS) onboard the Hubble Space Telescope (HST), we have identified a sample of RR Lyrae candidates in both dTrans galaxies [219 RRab (RR0) and 13 RRc (RR1) variables in ESO294-G010; 225 RRab and 44 RRc stars in ESO410-G005]. The metallicities of the individual RRab stars are calculated via the period-amplitude-[Fe/H] relation derived by Alcock et al. This yields mean metallicities of $\langle [Fe/H] \rangle_{ESO294} = -1.77 \pm 0.03$ and $\langle [Fe/H] \rangle_{ESO410} = -1.64 \pm 0.03$. The RRL metallicity distribution functions (MDFs) are investigated further via simple chemical evolution models; these reveal the relics of the early chemical enrichment processes for these two dTrans galaxies. In the case of both galaxies, the shapes of the RRL MDFs are well-described by pre-enrichment models. This suggests two possible channels for the early chemical evolution for these Sculptor group dTrans galaxies: 1) The ancient stellar populations of our target dwarf galaxies might have formed from the star forming gas which was already enriched through ``prompt initial enrichment'' or an ``initial nucleosynthetic spike'' from the very first massive stars, or 2) this pre-enrichment state might have been achieved by the end products from more evolved systems of their nearest neighbor, NGC 55. We also study the environmental effects of the formation and evolution of our target dTrans galaxies by comparing their properties with those of 79 volume limited ($D_{\odot} < $2 Mpc) dwarf galaxy samples in terms of the luminosity-metallicity relation and the {{\sc H~i}\/} gas content. The presence of these RRL stars strongly supports the idea that although the Sculptor Group galaxies have a considerably different environment from the Local Group (e.g. no giant host galaxies, loosely bound and very low local density), they share a common epoch of early star formation with the dwarf satellite galaxies in the Local Group. 
\end{abstract}


\section{Introduction}
The chemical evolution of galaxies during their earliest stages provides important insights for the initial physical and environmental conditions of the galaxy formation process. The star formation history (SFH) of galaxies is one of the important factors determining the path of their chemical evolution because the number of stars formed (i.e. the star formation rate) with a given initial mass function (IMF) controls the rate of chemical enrichment (Matteucci 2012). One key observable signature of the SFH of a galaxy is the metallicity distribution function (MDF) of its stars. Thus, we can test possible scenarios for the early epochs of galaxy formation by comparing the observed MDF of the oldest stellar population in the galaxy with several analytical models of chemical evolution. 

In this context, the Sculptor group dwarf galaxies provide an excellent test bed for investigating the early chemical evolution of dwarf satellite galaxies formed in low density environments (i.e. outskirts of dense galaxy clusters,  Local Group-like group environment, or field-like environment; see Section 4.4 of the present study). The Sculptor group, also known as the Sculptor filament, is one of the closest groups of galaxies to our Local Group (i.e. distance to the group center, $d_{gc} \sim $3.9 Mpc; Karachentsev et al. 2003). Five bright late-type galaxies (NGC 55, 247, 253, 300, and 7793) and at least 16 dwarf satellites form a very loosely bound, field-like system stretched along the line of sight over $\sim$5 Mpc. Our target galaxies, ESO294-G010 and ESO410-G005, are Phoenix-like transition-type dwarf galaxies (dTrans) that morphologically resemble the gas-poor dwarf spheroidal galaxies (dSphs) but contain detectable amounts of neutral hydrogen (H I) gas (i.e. $\sim10^5 M_{\odot}$; Bouchard et al. 2005; Jerjen, H., Freeman, K. C., \& Binggeli, B. 1998), and show signs of recent star formation activity. Most recently, from their detailed analysis of the color-magnitude diagrams obtained using deep HST/ACS imaging for the five Sculptor group dwarf galaxies, Lianou et al. (2013; hereafter L13) found population gradients in ESO294-G010 and ESO410-G005 in the sense that young blue main sequence stars (age $<$ 100 Myr) and intermediate age asymptotic giant branch stars (age $\sim$ 1-2 Gyr) are more centrally concentrated, while old horizontal branch stars (age $>$ 10 Gyr) tend to be more spatially extended. 

The high quality of those HST/ACS images also allowed the first detection of a significant population of RR Lyrae variable star candidates beyond the Local Group. Da Costa et al. (2010) were able to discover numerous RR Lyrae candidates in both dTrans galaxies using a template light curve technique (Layden 1998). The presence of the RR Lyrae stars in ESO294-G010 and ESO410-G005 directly confirms the existence of ancient stellar populations with ages $>$ 10 Gyr and indicates that the Sculptor group dwarf galaxies share a common epoch of the earliest star formation with our Local Group galaxies even though these two groups of galaxies have quite different local density environments. 

In addition to being a direct probe of ancient stellar populations, RR Lyrae stars provide a variety of utilities to investigate a number of important astrophysical applications; these stars are well-known reliable population II distance indicators. There is a correlation between the metallicities and the pulsation properties (i.e. periods and amplitudes) of the fundamental mode RR Lyrae stars (RR0 or RRab). Using this relation (Alcock et al. 2000; Sarajedini et al. 2006), we can calculate the metallicities of individual RRab stars, and thus construct the MDF for a purely old stellar population. Lastly, the intrinsic colors of RRab stars at their faintest luminosity are largely independent of their other physical properties [$(V-I)_{0,min}=$0.58$\pm$0.02; Guldenschuh et al. 2005], therefore using their observed colors at minimum light we can also estimate the line-of-sight reddening.       

The main goal of the present study is to examine the MDFs of legitimate RR Lyrae stars in the two Sculptor group dTrans galaxies, ESO294-G010 and ESO410-G005, in order to investigate the early chemical evolution of these systems, especially in light of the low density environments in which they reside. This paper is organized as follows. Section 2 provides a description of the data set and photometry process. In section 3 we report the results of our analysis including the general trends in the color-magnitude diagrams (CMDs) of the galaxies (Sec 3.1), the details of our RR Lyrae detection method and the pulsation properties of the RR Lyrae candidates (sec 3.2), the calculation of the metallicities of individual RR Lyraes (sec 3.3), and the distance measurements (sec 3.4). In section 4 based on our estimates in the previous sections, we present our in depth discussion on the star formation histories, early chemical evolutions, the luminosity-metallicity relations of the our target dTrans, and the environmental effects on the evolution of the near field dwarf galaxies. Finally in section 5 we present a summary of our results. 
    
\vskip 1cm

\section{Observations and Data Reduction}
The science images of ESO294-G010 and ESO410-G005 used in this study were taken with the Advanced Camera for Surveys Wide Field Channel (ACS/WFC) on board $Hubble$ $Space$ $Telescope$ ($HST$) as a part of GO-10503 (PI: Da Costa). The observing log for the retrieved archival images is shown in Table 1. The detailed observing strategy has been provided by Da Costa et al. (2010). To briefly describe the observations, the central regions of each galaxy were placed on one of WFC chips so that the imaging covers almost the entire visual extent of the targets. ESO294-G010 was imaged 12 times in the F606W and 24 times in the F814W filter with an exposure time of 1160s during the observing baseline of $\sim$ 1.86 days. The same number of exposures were obtained for the observations of ESO410-G005 with an exposure time of 1120s over the observing baseline of 4.55 days. In this way the observations provide a good quality time-series photometry that allows us to detect genuine RR Lyrae variable candidates in each galaxy. 
 
\begin{deluxetable*}{ccccccc}
\tabletypesize{\scriptsize}
%\rotate
\tablecaption{Observation Log. \label{tbl-1}}
\tablewidth{0pt}
\tablehead{
\colhead{Object}  & \colhead{RA (J2000)} & \colhead{Dec. (J2000)} & 
\colhead{Filters} & \colhead{Data Set}   & \colhead{HJD Range (+2 453 000)} 
}
\startdata
ESO294-G010 & 00 26 33.40   & -41 51 19.0  & F606W   & 12x1160s  & j9bd0701*,j9bd0801*,j9bd1202* & 859.27358 -- 861.13659 \\
           &               &              & F814W   & 24x1160s  & j9bd0702*,j9bd0802*,j9bd0901*--j9bd1201* & 859.40613 -- 861.00342\\ 
           &               &              &         &           &                                          &                       \\
ESO410-G005 & 00 15 31.40   & -32 10 47.0  & F606W   & 12x1120s  & j9bd1301*,j9bd1401*,j9bd1802* & 861.40425 -- 863.21702 \\
           &               &              & F814W   & 24x1120s  & j9bd1302*,j9bd1402*,j9bd1501*--j9bd1801* & 861.53711 -- 866.08158\\     
\enddata
\end{deluxetable*}

We photometered the point sources in the cosmic-ray cleaned \_$crj$ images with the ACS module of DOLPHOT package (Dolphin 2000) by following the standard procedures for DOLPHOT. These \_$crj$ images are calibrated FITS images that were produced through the ACS Calibration (CALACS) pipeline, which performs basic image reduction and  data calibration including cosmic ray rejection. Pre-constructed point spread functions (PSFs) for each ACS passband were used to expedite the PSF photometry. Once bad pixels were removed, each ACS/WFC image which contains both chips was split into two separate image files with single-chip format. Then, all of the individual images were aligned and photometered simultaneously. Aperture corrections were calculated for each image with the default settings of DOLPHOT. The instrumental magnitudes were transformed to the native ACS/WFC VEGAmag system as well as the ground-based Johnson-Cousins VI system using the calibration equations of Sirianni et al. (2005) implemented in DOLPHOT. The resultant magnitudes were corrected for the loss of charge transfer efficiency (CTE) as described in ACS ISR 2003-09. We performed photometric completeness tests using the artificial star feature (``acsfakelist'') in DOLPHOT. This experiment allows us to gauge the degree of photometric completeness, as well as photometric errors. 

The final set of standard VI photometry was extracted from the resultant file by selecting only those stars with ``object type'' equal to 1 (i.e. a good star). These stars also meet the additional selection criteria set by the several photometric quality parameters: $S/N >$ 10, --0.5 $< Sharpness_{F606W} <$ +0.5, and --1 $< Crowding_{F606W} <$ +1. The sharpness approaches zero for a star with a nearly perfect stellar profile, while for a source that is too sharp (i.e. a cosmic ray) this value becomes positive. A negative value indicates a source that is extended or too broad (i.e. a blend, cluster or galaxy). According to the test done by Dolphin et al. (2004), good stars in an uncrowded field should have sharpness values between -0.3 or +0.3. The crowding parameter is defined in magnitude units. For an isolated star, the value is zero. For stars in a highly crowded field, this value becomes larger than unity (positive or negative).

\section{Results}
\subsection{Color-Magnitude Diagrams}
Color-magnitude diagrams (CMDs) of ESO294-G010 and ESO410-G005 in the standard VI system are presented in Figures 1 and 2.  Our artificial star tests show that the point sources in both galaxies are well photometered to $\sim$ 2 magnitude below the horizontal branch (HB) level with better than $\sim$ 50\% photometric completeness (see the right panels of Fig 1 \& Fig 2). At the level of the HB magnitude the photometric completeness reaches $\sim$ 88\%. Therefore we assume that photometric incompleteness does not significantly affect our analysis of the RR Lyrae variable stars. 

\begin{figure}
\epsscale{1.3}
\plotone{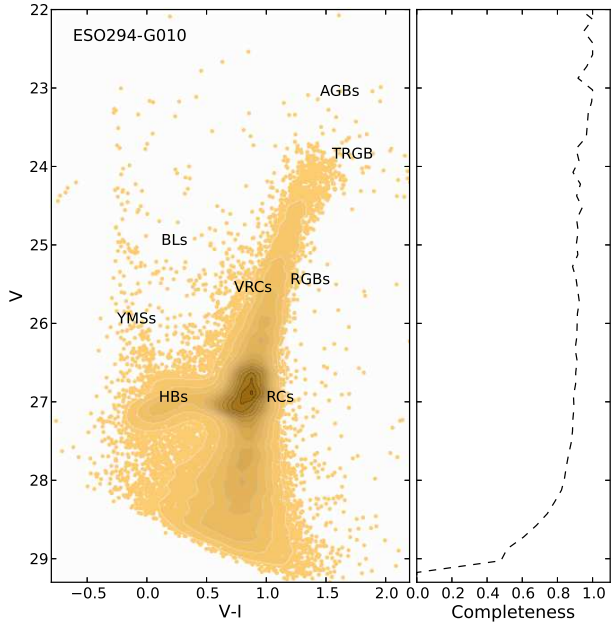}
\caption{The VI color-magnitude diagram of ESO294-G010 (see section 3.1. for the detail explanations of each label). The right-hand panel illustrates the photometric completeness indicating that our photometry is as complete as $\sim$ 88 \% at the level of horizontal branch magnitude.\label{fig1}}
\end{figure}

\begin{figure}
\epsscale{1.3}
\plotone{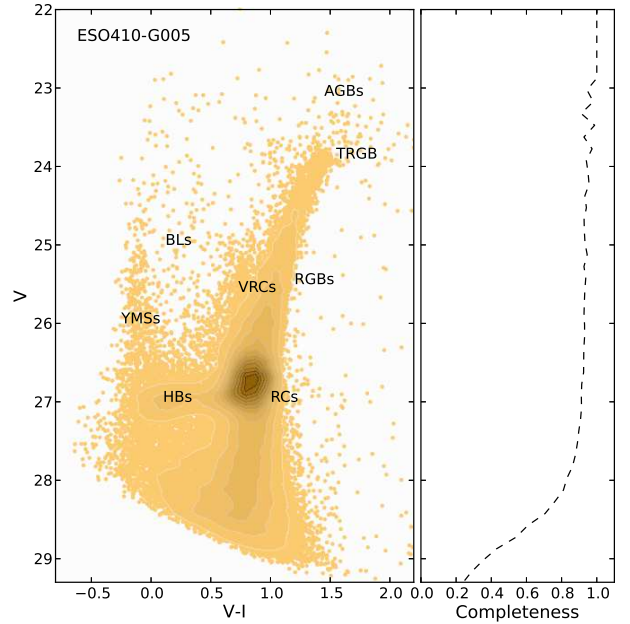}
\caption{The VI color-magnitude diagram of ESO410-G005 (see section 3.1. for the detail explanations of each label). The photometric completeness is comparable to that for ESO294-G010.\label{fig2}}
\end{figure}

According to the morphological type $T$ (de Vaucouleurs et al. 1991; Karachentsev et al. 2004), and the classifications for dwarf galaxies (Mateo 1998), both ESO294-G010 ($T$=--3) and ESO410-G005 ($T$=--1) satisfy the criteria of transition-type dwarf galaxies (dTrans) that exhibit reduced recent star formation but relatively high gas fractions. The tidal indices ($\Theta$; Karachentsev et al. 2004, 2013; also see Section 4.4. of the present study) of these two dTrans systems indicate that they are gravitationally bound to their nearest neighbor, NGC 55 which is a nearly edge-on barred irregular galaxy. 

In general, given that these two dTrans galaxies have comparable mass and size, the CMDs of ESO294-G010 and ESO410-G005 show a close resemblance to each other. Both of the two dTrans feature diverse stellar constituents with wide ranges of age and metallicity: young blue main-sequence stars (YMSs), relatively weak but recognizable intermediate mass He core-burning blue loop stars (BLs), a handful of vertical extension red clump stars (VRCs) and asymptotic giant branch stars (AGBs) both of which represent intermediate-age (1 -- 10 Gyr) populations, dominant first ascent red giant branch stars (RGBs), tightly gathered red clump stars (RCs), and distinctively stretched horizontal branch stars (HBs) that strongly signal an ancient stellar population (age$>$10 Gyr). Detailed descriptions of each characteristic stellar constituent in these two galaxies can be found in the recent work of L13.   

\subsection{RR Lyrae Stars}
\subsubsection{Detection and Characterization}
We employed the technique of Yang et al. (2010; Y10 hereafter) and Yang \& Sarajedini (2012; YS12 hereafter) to identify and characterize RR Lyrae stars in ESO294-G010 and ESO410-G005. In summary, we start by searching all possible variable star candidates at the level of the HB brightness (26.5 $<$ V $<$ 27.6). The variability of each star within the given range of V magnitude was evaluated by the reduced $\chi^2_{VI}$ defined by the following formula: 

\begin{displaymath}
\chi^2_{VI} = \frac{1}{N_V + N_I} \times
\Bigg[\sum_{i=1}^{N_V} \frac{(V_i - \overline V)^2}{\sigma_i^2} +  
\sum_{i=1}^{N_I} \frac{(I_i - \overline I)^2}{\sigma_i^2}\Bigg], (1)
\end{displaymath}.

Any data points located more than $\pm$3$\sigma$ from the mean magnitude were excluded from the calculation of $\chi^2_{VI}$. Only those stars with $\chi^2_{VI}$ values greater than 3.0 were considered as potential variable candidates [i.e. typical non-variable stars at the HB level (V(HB)$\sim$27) have $\chi^2_{VI}$ values less than 3.0]. The VI time series photometry of the variable star candidates that satisfy our variability criterion was analyzed with our template light curve fitting routine, RRFIT, which fits both V and I band data simultaneously yields the best fitting light curve parameters such as period, amplitude, epoch at maximum light, and mean magnitudes (see YS12 for more details). Each output best-fit light curve was carefully examined by eye. Then in the VI CMDs, we checked the mean color and magnitude of those RR Lyrae candidates that passed our eye-inspection to make sure they are located in the expected region of the HB.  Finally, in order to further ensure that our sample contains genuine RR Lyrae variables, we applied marginal V-I color ranges using the blue and red edges of the empirical instability strip for RR Lyrae variables from Mackey \& Gilmore (2003) to which we added a 3$\sigma$ $rms$ error (i.e. $\sigma_{VI} = \pm 0.04$ mag) in the (V-I) colors at the level of the HBs. Figure 3 illustrates zoomed-in CMDs with the RR Lyrae candidates identified from our analysis. Both CMDs show that the bulk of genuine RRL candidates are well selected within the adopted V-I color ranges for each target galaxy. 

\begin{figure}
\epsscale{1.2}
\plotone{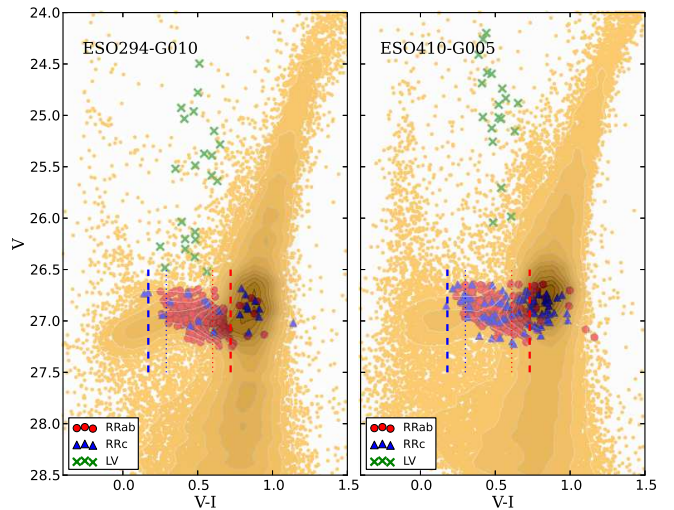}
\caption{The zoomed-in VI color-magnitude diagrams with RR Lyrae candidates shown. The dotted lines illustrate the blue and red edges of the empirical instability strip (Mackey \& Gilmore 2003) for each galaxy, while the dashed lines which were extended to both blue and red sides of the empirical instability strip by 3$\sigma_{VI}$ indicate the V-I color range for our selection of the RR Lyrae candidates. Doing so we increase genuine RR Lyrae samples in each dTran as many as possible; The green crosses indicate the luminous variable (LV) candidates. These stars probably include Anomalous Cepheids, Pop II Cepheids, and short-period Classical Cepheids.\label{fig3}}
\end{figure}

After applying the above procedures, we have identified and characterized 232 RR Lyrae stars (219 RRab and 13 RRc) in ESO294-G010 and 269 RR Lyraes (225 RRab and 44 RRc) in ESO410-G005. Now we go back to the reference images to examine the individual RR Lyraes to see how many blended objects were included among the selected RR Lyrae samples. Our examination reveals that there are 73 blended objects (68 RRab + 5 RRc) among the 232 RRL samples in ESO294-G010, while we found 92 blended objects (73 RRab + 19 RRc) among the 269 RRL candidates in ESO410-G050. We further examine the pulsation properties (see Fig 9. Period-amplitude diagrams in section 3.2.3) of these objects to investigate whether the blending causes any systematic bias in the mean pulsation properties of our RR Lyrae samples. We find no indication of significant systematic biases in the period-amplitude space due to the blending; instead, we see that the blended objects follow the mean period-amplitude trend just like the other well-isolated RR Lyrae stars. Therefore we assume that our analyses in this study are not negatively affected by the effects of stellar blending. Figures 4--5 illustrate a sample of the best-fitting light curves for the RR Lyrae candidates. The basic properties of each of these RR Lyrae candidates are summarized in Tables 2-3.  

\begin{figure}
\epsscale{1.4}
\plotone{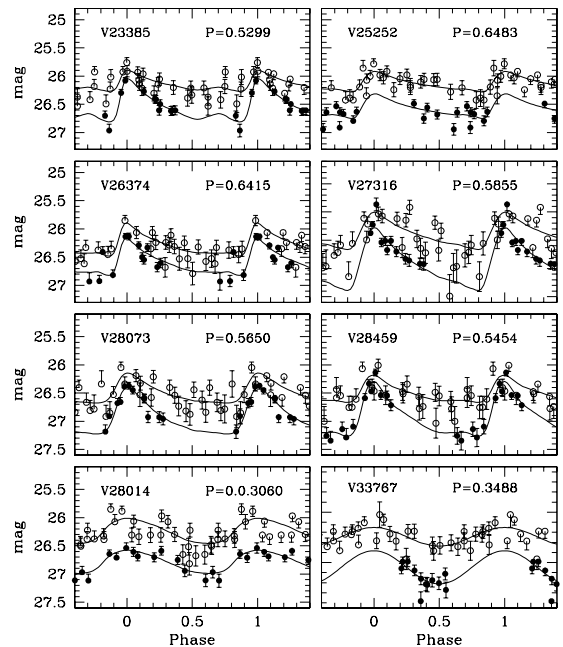}
\caption{A sample of the best-fitting light curves for the RR Lyrae candidates in ESO294-G010. The {\it solid dots} represent V-band data, while I-band data are marked by {\it open circles}.\label{fig4}}
\end{figure}

\begin{figure}
\epsscale{1.4}
\plotone{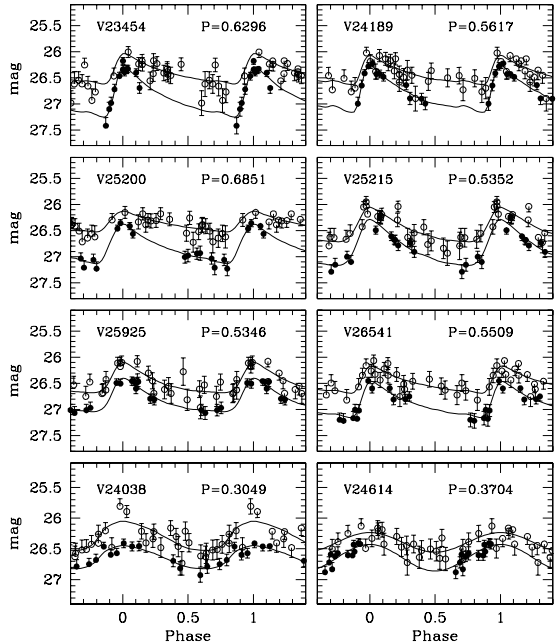}
\caption{Same as Figure 4, but for ESO410-G005.\label{fig5}}
\end{figure}

We also attempted to search for other types of variable stars in both dTrans by expanding our search ranges in the color-magnitude ($24.0 < V < 26.7$; $-0.5 < V-I < 0.95$) and period ($0.4 < P < 3$ days) spaces. Then we apply the same method as we did for the RR Lyrae survey. We have uncovered a significant number of luminous variable (LV) candidates in both dTrans (22 in ESO294-G010; 19 in ESO410-G005). Their photometric and pulsation properties are summarized in Table 4. Figure 6 illustrates some examples of the best-fitting light curves for the LV candidates in the two dTrans. These stars are brighter than the RRL group by 0.5 $\sim$ 3.0 mag (see Fig 3). Their color-magnitude positions and period-amplitude properties appear to be consistent with those of typical Anomalous Cepheids, Pop II Cepheids, and short-period Classical Cepheids (Gallart et al. 2004; Pritzl et al. 2005; Bernard et al. 2009). We find that these LVs are more concentrated in the inner fields and their spatial distributions seem to be aligned with the major axis of the ellipsoid of each dTran, which might indicate that these stars belong to the young or intermediate-age populations of the dTrans (see the bottom panels of Fig 12).

\begin{figure}
\epsscale{1.4}
\plotone{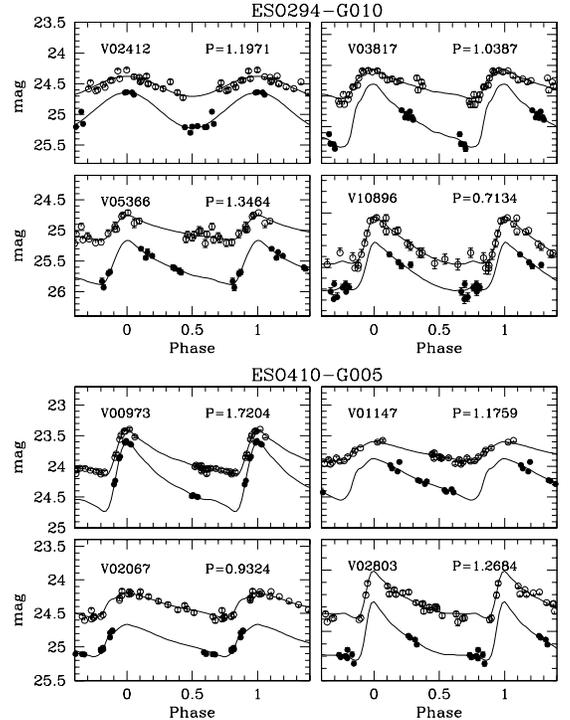}
\caption{A sample of the best-fitting light curves for the luminous variable candidates in our target galaxies. The {\it solid dots} represent V-band data, while I-band data are marked by {\it open circles}.\label{fig4}}
\end{figure}

\subsubsection{Synthetic Light Curve Simulation}
The derived pulsation parameters of RR Lyrae variables are often subject to spurious periods (or aliases) that are mainly caused by the effects of undersampling. In order to estimate the degree of any aliases in our period-searching analysis, we have performed extensive simulations and statistical tests (Y10; YS12; Sarajedini et al. 2012). Input periods and amplitudes were randomly generated within an appropriate range of ab-type and c-type RR Lyraes. We then produced $\sim$1000 synthetic RR Lyrae light curves with these input periods and amplitudes by applying the observing windows - such as observing baseline, number of epochs, cadence, and photometric errors - that were extracted from the time series photometry of a well measured RR Lyrae candidate. Finally these synthetic VI time series photometry sets were analyzed by the RRFIT routine to check how well the output periods and amplitudes recover the input values. The comparison between the input and output solutions is presented in Figures 7--8 where we plot the distributions of the input and output periods ($\Delta$P), and the difference between these two values as a function of input period. 

\begin{figure}
\epsscale{1.3}
\plotone{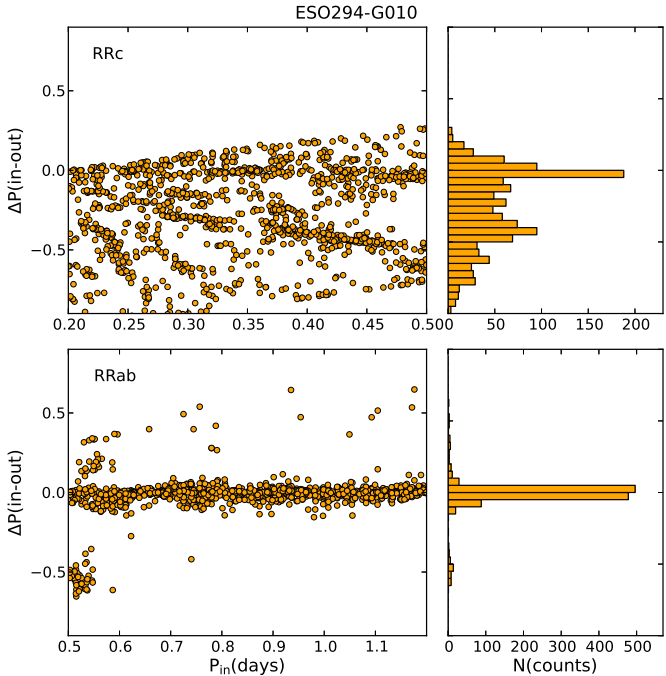}
\caption{The results of our synthetic light curve simulations for ESO294-G010. Left panels illustrate the difference ($\Delta P=P_{in}-P_{out}$) between the input and output period as a function of the input period for two different pulsation modes. Right panels show the $\Delta P$ distributions. Our simulations reveal that type-c RR Lyrae stars are significantly influenced by aliasing while type-ab RR Lyrae stars are almost free from the effects of aliasing. \label{fig7}}
\end{figure}

\begin{figure}
\epsscale{1.3}
\plotone{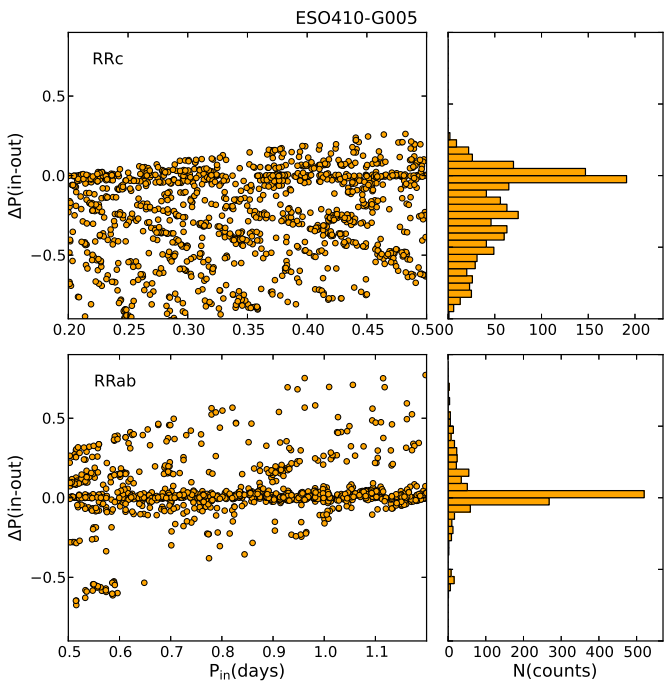}
\caption{Same as Figure 7, but for ESO410-G005. Similar to the case of ESO294-G010, type-c RR Lyrae stars in ESO410-G005 are more severely affected by aliasing as compared with type-ab RR Lyrae stars.\label{fig8}}
\end{figure}

The next step involved the following statistical test. From the artificial RR Lyrae lists, we randomly sample artificial RR Lyrae stars with the same number as the observationally detected RR Lyraes in each galaxy [ESO294-G010 : 219 (RRab) and 13 (RRc); ESO410-G005 : 225 (RRab) and 44 (RRc)] to examine their $\Delta$P distribution. The 1-$\sigma$ error ($\sigma_{P}$) of this $\Delta$P distribution from the best-fit Gaussian can be considered as a good estimate of the standard error in the individual periods derived by our period-searching analysis. In order to improve the statistical robustness of the $\sigma_{P}$ value, this random sampling was iterated 10,000 times. We then consider the average spread ($<\sigma_{P}>$) of these 10,000 samples as a realistic error for our period measurements. The same approach was applied to estimate errors in the V-band amplitude. 

We summarize here the main findings of our simulations. 
\begin{enumerate}
\item For the RRab stars in ESO294-G010, $\sim$ 80\% of the input periods were well recovered within $\pm$ 0.05 day, while this recovery fraction was significantly reduced to $\sim$ 25\% for the RRc stars. In the case of ESO410-G005, this value becomes $\sim$ 70\% and 30\% for the RRab and RRc respectively.  

\item The standard errors ($\sigma_{P}/\sqrt{N}$) of the individual RR Lyrae periods in our analysis are $\sigma(P_c)$=$\pm$ 0.0214 and $\sigma(P_{ab})$=$\pm$ 0.0006 days for the RR Lyrae candidates found in ESO294-G010. For the ESO410-G005 RR Lyrae candidates, we find $\sigma(P_c)$=$\pm$ 0.0061 and $\sigma(P_{ab})$=$\pm$ 0.0008 days. The errors in the V band amplitudes are $\sigma(Amp(V)_{c})$ = 0.0214 mag and $\sigma(Amp(V)_{ab})$ = 0.0016 mag for the ESO294-G010 RR Lyraes, while these values become $\sigma(Amp(V)_{c})$ = 0.0013 mag and $\sigma(Amp(V)_{ab})$ = 0.0018 mag for the ESO410-G005 RR Lyraes.

\item We found no significant biases in our determination of the mean periods and amplitudes for the RRab stars. However, it is evident for both galaxies that RRc stars are subject to more significant aliasing as compared to RRab stars. In addition there exists a systematic trend in the sense that the output periods of RRc stars tend to be longer than the input periods. Thus we conclude that the aliasing in the periods of RRc stars significantly hinders the accurate determination of the pulsation periods of these stars, and it is not possible in this study to utilize their pulsation properties to study other aspects of RR Lyrae stars.

\end{enumerate}

\subsubsection{Period-Amplitude Diagrams}
It is well known that the period-amplitude (P-A) relations (i.e. Bailey diagram) of RR Lyrae stars in the Milky Way Galaxy (both in the Galactic Globular Clusters and the field Galactic Halo) exhibit the so called the Oosterhoff dichotomy: Oosterhoff type I (Oo I) systems tend to have intermediate metallicities and a mean pulsation period of $\sim$ 0.55 days in the fundamental mode with a low RRc frequency ($N_{c} = n_{c} /n_{abc} =0.2$), while Oosterhoff type II (Oo II) systems are generally metal-poor and have mean pulsation periods of $\sim$ 0.65 days with a higher fraction of RRc stars ($N_{c} =0.45$; Castellani \& Quarta 1987). Thus the Bailey diagram can serve as a useful diagnostic tool for investigating the basic pulsation properties of the RR Lyrae populations.

\begin{figure}
\epsscale{1.3}
\plotone{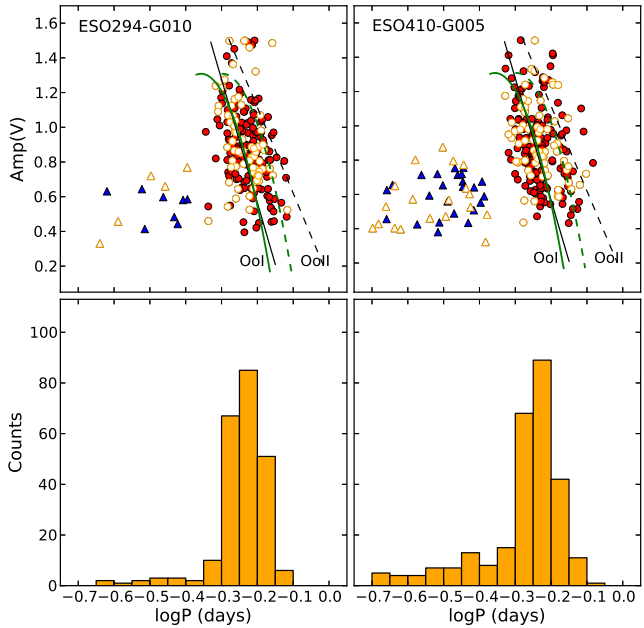}
\caption{The Period-Amplitude (P-A) relations of the RR Lyrae candidates in ESO294-G010 and ESO410-G005 are plotted in the top panels. The bottom panels show the period distributions. The open symbols[circles (RRab or RR0); triangles (RRc or RR1)] represent the blend objects (see section 3.2.1), while the well-isolated RR Lyrae samples are marked with filled symbols [{\it blue triangles} and {\it red dots}]. Solid and dashed lines indicate the fiducials of Oosterhoff I and II systems [straight lines (Clement \& Rowe 2000); quadratic relations (Cacciari et al. 2005, Zorotovic et al. 2010)]. The pulsation properties of the RR Lyrae candidates in these two Sculptor group dTrans appear to follow the typical trend for the RR Lyrae populations of Local Group dwarf satellite galaxies.\label{fig9}}
\end{figure}

In Figure 9 we plot the P-A relations and period distributions of our RR Lyrae candidates found in ESO294-G010 and ESO410-G005. The loci of Oosterhoff type I and II for the Galactic Globular Cluster (GGC) systems [straight lines (Clement \& Rowe 2000); quadratic relations (Cacciari et al. 2005, Zorotovic et al. 2010)] are also plotted for comparison. Solid symbols [red dots (RRab); blue triangles (RRc)] indicate the well-isolated RR Lyraes stars not affected by crowding, while the blended objects (see section 3.2.1) are marked with open symbols. According to Figure 9, the blended stars are  indistinguishable from the other well-isolated RR Lyrae variables. Thus, we find no significant systematic biases in the period-amplitude space due to stellar blending. We see that the bulk of the RRab candidates in these two dTrans galaxies appear to be more tightly distributed around the locus of the Oosterhoff I type GGCs, while there are no significant population of Oosterhoff II RR Lyraes in both galaxies. The mean pulsation periods of the RRab candidates are $\langle P \rangle_{ab,ESO294}$ = 0.5932 $\pm$ 0.0003 (error1) $\pm$ 0.0006 (error2) days and $\langle P \rangle_{ab,ESO410}$ = 0.5898 $\pm$ 0.0003 (error1) $\pm$ 0.0008 (error2) for ESO294-G010 and ESO410-G005 respectively. The quoted value of ``error1'' represents the standard error of the mean and the ``error2'' value is the error derived from our synthetic light curve simulations. 

\subsection{Metallicity}
RR Lyrae stars are very useful for tracing the early chemical enrichment histories of dwarf satellite galaxies. These pulsating variables represent a purely old stellar population (age $>$ 10 Gyr) in a given galaxy. Hence, the advantage of using RR Lyraes for a robust measurement of the metal abundance of an ancient stellar population particularly stands out in the case of studying dTrans galaxies which exhibit recent star formation and diverse stellar populations with wide ranges of ages and metallicities.  

The metal abundance of ab-type RR Lyraes can be calculated using the correlation between period, amplitude and $[Fe/H]$ given by Alcock et al. (2000) in the following form

\begin{displaymath}
[Fe/H] = -8.85 [log P_{ab} + 0.15 Amp(V)] -2.60,      (2)
\end{displaymath}

where Amp(V) indicates the V band amplitude, and [Fe/H] is in the Zinn \& West (1984) metallicity scale. 

\begin{figure}
\epsscale{1.0}
\plotone{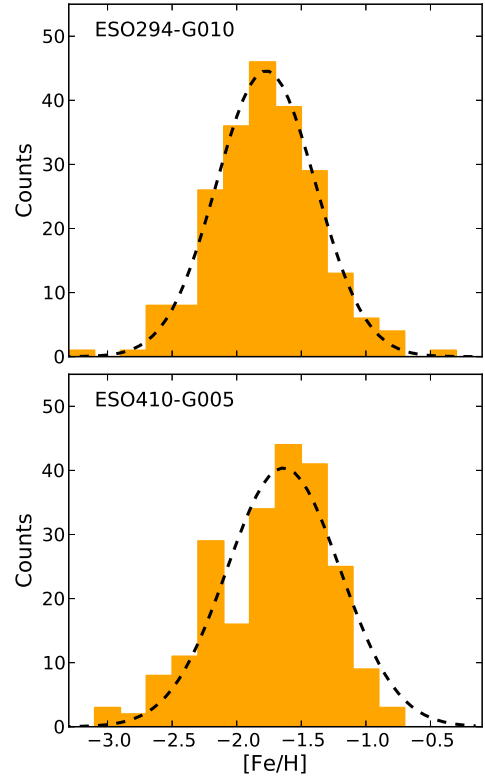}
\caption{The metallicity distribution functions (MDFs) of RRL candidates in ESO294-G010 and ESO410-G005. The dashed lines indicate a Gaussian fit to these data. The best-fit Gaussians yield a mean of $[Fe/H]=-$1.77 and a standard deviation of 0.38 dex for ESO294-G010, and for ESO410-G005 these value become $[Fe/H]=-$1.64 and 0.44 dex.\label{fig10}}
\end{figure}

Figure 10 shows the metallicity distribution functions (MDFs; binned histograms and the best-fit gaussians) of the RRab stars found in the two dTrans galaxies of the present study. The MDF of the ESO294-G010 RRab stars exhibits a slightly lower metallicity peak as compared to that of  the ESO410-G005 RRab stars. The mean metallicities estimated from the best-fit Gaussians to the MDFs are $\langle [Fe/H] \rangle=$--1.77 $\pm$ 0.03 (sem) for ESO294-G010, and, $\langle [Fe/H] \rangle=$-1.64 $\pm$ 0.03 (sem) for ESO410-G005 respectively. The uncertainties shown as sem here represent the standard errors of the mean values. The systematic errors are likely larger and closer to $\sim$0.2 dex as demonstrated by Jeffery et al. (2011).

To better understand the properties of the RR Lyrae populations in these two dTrans galaxies, we now compare the RR Lyrae MDFs with those of the RGB stars. We construct MDFs of the RGB stars for both ESO294-G010 and ESO410-G005 by interpolating (Sarajedini \& Jablonka 2005) within a set of Dartmouth isochrones (Dotter et al. 2008) that were generated to have metallicity values ranging from -2.5 to +0.5 dex at a fixed age of 12.5 Gyr. We used three values of $[\alpha/Fe]=$0.0, 0.2, and 0.4 dex to gauge the effects of $\alpha$ element enhancement on the overall shape of the RGB MDF. For each galaxy, the RGB stars were selected within a magnitude range of $I_{TRGB} < I < 25 $ (i.e. $I_{TRGB,ESO294}=$22.33; $I_{TRGB,ESO410}=$22.36, see the next section for the details). To account for interstellar extinction, reddening values of E(B-V) = 0.006 and 0.014 for ESO294-G010 and ESO410-G005 were adopted respectively from the Galactic extinction map of Schlegel, Finkbeiner \& Davis (1998).     

\begin{figure*}
\epsscale{1.0}
\plotone{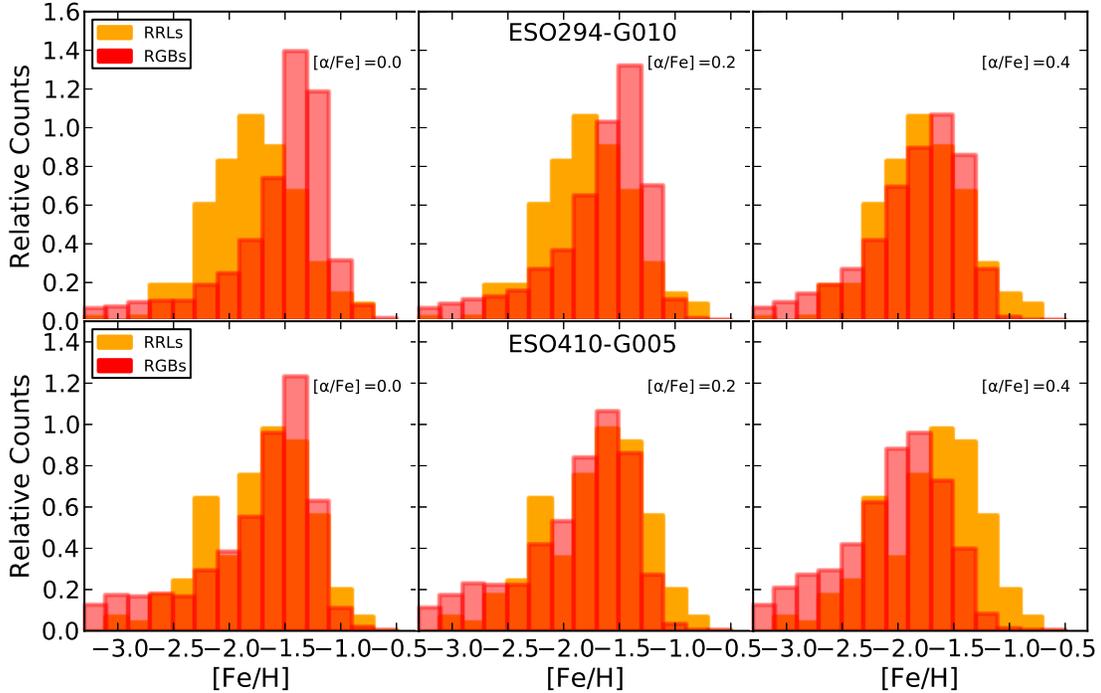}
\caption{$[Fe/H]$ distributions for the RRab (orange) and RGB (red) stars in ESO294-G010 and ESO410-G005. The MDFs of the RGB stars were constructed using interpolation (Sarajedini \& Jablonka 2005) wihtin a grid of Darthmouth isochrones with metallicity values ranging from -2.5 to +0.5 dex at a fixed age of 12.5 Gyr.\label{fig11}}
\end{figure*}

Figure 11 illustrates the resulting MDFs of the RGB stars compared to those of the RR Lyrae stars. The MDFs are normalized to have the same total counts. In general, the overall shapes of the resultant RGB MDFs are not Gaussian but tend to be skewed toward lower abundances. We also see that the MDFs of the RGB stars are affected by the assumed values of $\alpha$ element ratio in the sense that higher the $\alpha$ element enhancement, the lower the overall metallicity. For ESO294-G010, the RRL stars appear to have a lower mean metal abundance as compared to the RGB stars at $[\alpha/Fe]=$0.0 and 0.2 dex while the MDFs of these two stellar populations agree fairly well at $[\alpha/Fe]=$0.4. In contrast with ESO294-G010, the metallicities of the RRL and RGB stars in ESO410-G005 generally agree well with each other except for where $[\alpha/Fe]=$0.4 dex. 

As seen from the CMDs of these two galaxies, their RGB stars exhibit a broad range in $(V-I)$ colors and noticeable features of intermediate age (1 $<$ age $<$ 10 Gyr) populations. It is worth mentioning here that Rejkuba et al. (2011) demonstrate that using the locations of RGB stars relative to a grid of standard RGB sequences with different metallicities in order to estimate the metal abundance of a diverse stellar population can lead to systematic effects. In particular, a metallicity bias of 0.1 -- 0.2 dex in the metal-poor direction can be introduced by age shifts from 12 Gyr to 8 Gyr, and this can cause a metallicity distribution to be skewed to the metal-poor end. 

The comparisons presented above suggest that the RR Lyrae MDFs may be more reliable for studying the early chemical enrichment history of dwarf galaxies because RR Lyrae stars are not subject to the age-metallicity degeneracy, and the overall shape of the RRL MDF is not as significantly influenced by an assumed value of the $[\alpha/Fe]$ ratio as compared with MDFs generated using RGB stars. As a result, our estimates of the mean metallicities of type-ab RR Lyrae stars are free from the metal-poor bias induced by intermediate-age stellar populations; furthermore, the handful number of RR Lyrae candidates within the empirical instability strips in each target galaxy greatly reduces possible bias in the obtained metallicity values due to the evolutionary effects of these pulsation HB stars from the zero age HB (ZAHB). Therefore, we consider our metallicity estimates to be faithful to the purely ancient stellar populations in ESO294-G010 and ESO410-G005. 

Now we turn our attention to whether the metallicities of the RRab samples show any radial trends throughout our target dTrans. Previously a metallicity gradient has been reported for the RR Lyrae stars in Tucana, one of the Milky Way dTrans in the sense that the more luminous, longer period (i.e. more metal-poor) RR Lyrae variables tend to be located at larger distances from the galactic center as compared to the fainter, shorter period (i.e. more metal-rich) RR Lyraes (Bernard et al. 2008). This phenomenon indicates that such  a metallicity gradient should have occurred in a very early stage of evolution of this isolated dTran. 

We examine the spatial distributions of our RRab samples in the fields of ESO294-G010 and ESO410-G005 in order to see if any radial trends similar to the case of Tucana exist in these Sculptor group dTrans. To do this, we divide our RRab samples into three groups [i.e. metal-rich ($[Fe/H]  >  \langle [Fe/H] \rangle +0.2$ dex) , metal-poor ($[Fe/H] <  \langle [Fe/H] \rangle -0.2$ dex), and metal-intermediate ($\langle [Fe/H] \rangle -0.2 < [Fe/H] <  \langle [Fe/H] \rangle + 0.2$)] according to their metallicity ranges. Then we plot their locations on top of the elliptical isophote which corresponds to the visual extent of each dTran. From Figure 12, we are able to see that all three groups of RRab samples are well mixed with each other from the inner to the outer regions; therefore we conclude that there are no significant metallicity gradients among the RRab stars for both Sculptor Group dTrans. We also tested for a radial trend in the brightnesses of the RRab stars (see the bottom panels of Fig. 12). Similar to the previous test for metallicity, we do not see any systematic difference in their spatial distributions between luminous [$V < V(HB)$] and faint [$V > V(HB)$] groups. Further we checked the spatial distributions of the peculiar peak at around $-2.2$ dex (blue open circles) in the MDF of the ESO410-G005 RRab samples. As we can see from the figure, their spatial distributions do not show any peculiarity. 

\begin{figure*}
\epsscale{1.0}
\plotone{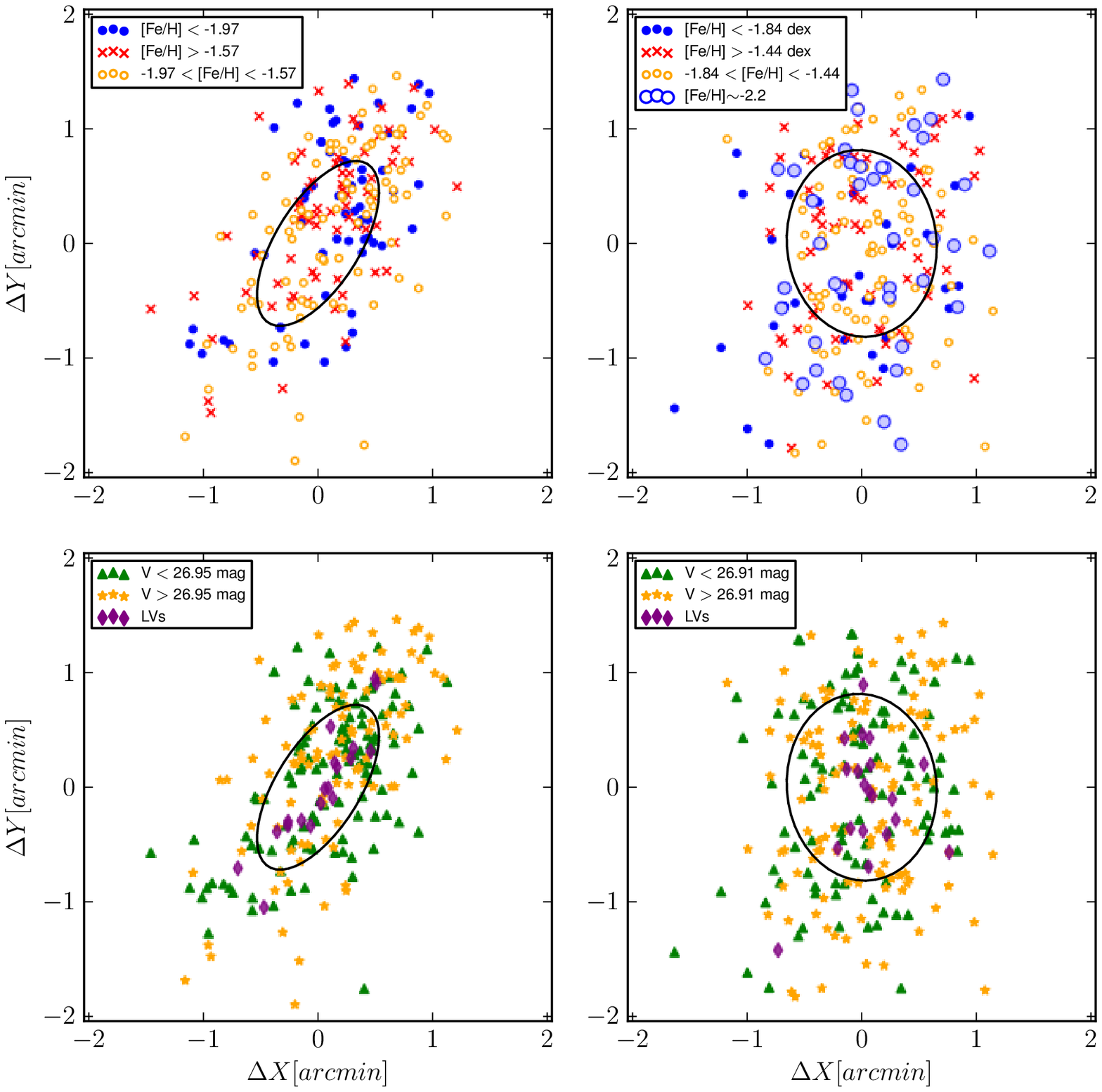}
\caption{The spatial distribution of the RRab stars in each dTran. The ellipses represent the isophotes which correspond to the visual extents of each galaxy. Top panels show the spatial trend of the metal-rich (red cross), metal-poor (blue dot), and metal-intermediate (orange circle) groups. Bottom panels illustrate the spatial distributions of the luminous (green triangle; $V < V(HB)$) and faint (orange star; $V > V(HB)$) groups.  The luminous variables (LVs; see section 3.2.1) are presented as $purple$ diamonds showing that these stars are more centrally concentrated and aligned with the major axis of their host galaxy as compared to the RR Lyrae stars. \label{fig12}}
\end{figure*}

Most recently L13 presented a very detailed analysis of the spatial distributions of several different stellar populations in different age groups in these two dTrans. Their analysis revealed that the old stellar populations such as HB stars are more uniformly distributed throughout the galaxies, while the younger populations such as young blue main sequence stars or AGB stars are more concentrated to the inner regions of the dTrans. The results from our analysis of the spatial distributions of the RRab samples in ESO294-G010 and ESO410-G005 are consistent with those of L13 indicating that the physical and chemical processes in the early evolutionary phase of these Sculptor Group dTran seem to be different from the case of the isolated Milky Way dTran, Tucana.

\subsection{Reddening and Distance}
We adopt reddening values of E(B-V) = 0.006 and 0.014 for ESO294-G010 and ESO410-G005, respectively from the Galactic extinction map of Schlegel, Finkbeiner \& Davis (1998) in the measurement of the distances to each galaxy. We are not able to estimate the line-of-sight reddenings toward the target galaxies using the minimum-light colors of the RRab stars because of the relatively low quality of the I-band light curves. 

Along with the metallicity values of the individual RR Lyrae stars estimated in the previous sections, we now have the foundation for the calculation of accurate distances of our target galaxies. The mean V magnitude of the 219 RRab stars detected in ESO294-G010 is $\langle V(RR) \rangle=$26.95$\pm$0.01 mag. The absolute V magnitudes of the individual RRab stars are calculated using the calibration given by Chaboyer (1999), $M_{V}= 0.23 [Fe/H] + 0.93$. This gives a mean absolute V magnitude of $\langle M_{V}(RR) \rangle=0.53 \pm 0.07$. The quoted error is the amount of uncertainty propagated from our measurement of the metal abundances of the RRab stars using the Alcock relation. By employing the extinction law, $A_{V}=3.1 E(B-V)=$0.019 we obtained a distance modulus of $(m-M)_{0}= 26.40 \pm 0.07$ mag which places ESO294-G010 at a distance of $\sim$ 1.9 $\pm$ 0.1 Mpc. 

For ESO410-G005, the mean V magnitude of 225 RRab stars is $\langle V(RR) \rangle=$26.91 $\pm$ 0.01 mag. Applying once again the calibration of Chaboyer (1999) for the luminosity-metallicity relation for RRab stars, and the line-of-sight extinction $A_{V}=$0.043, we obtain $\langle M_{V}(RR) \rangle=$0.53 $\pm$ 0.07 mag. This yields a distance modulus of $(m-M)_{0} =$ 26.33 $\pm$ 0.07 mag which is equivalent to a distance of $\sim$ 1.9 $\pm$ 0.1 Mpc. 

\begin{figure}
\epsscale{1.2}
\plotone{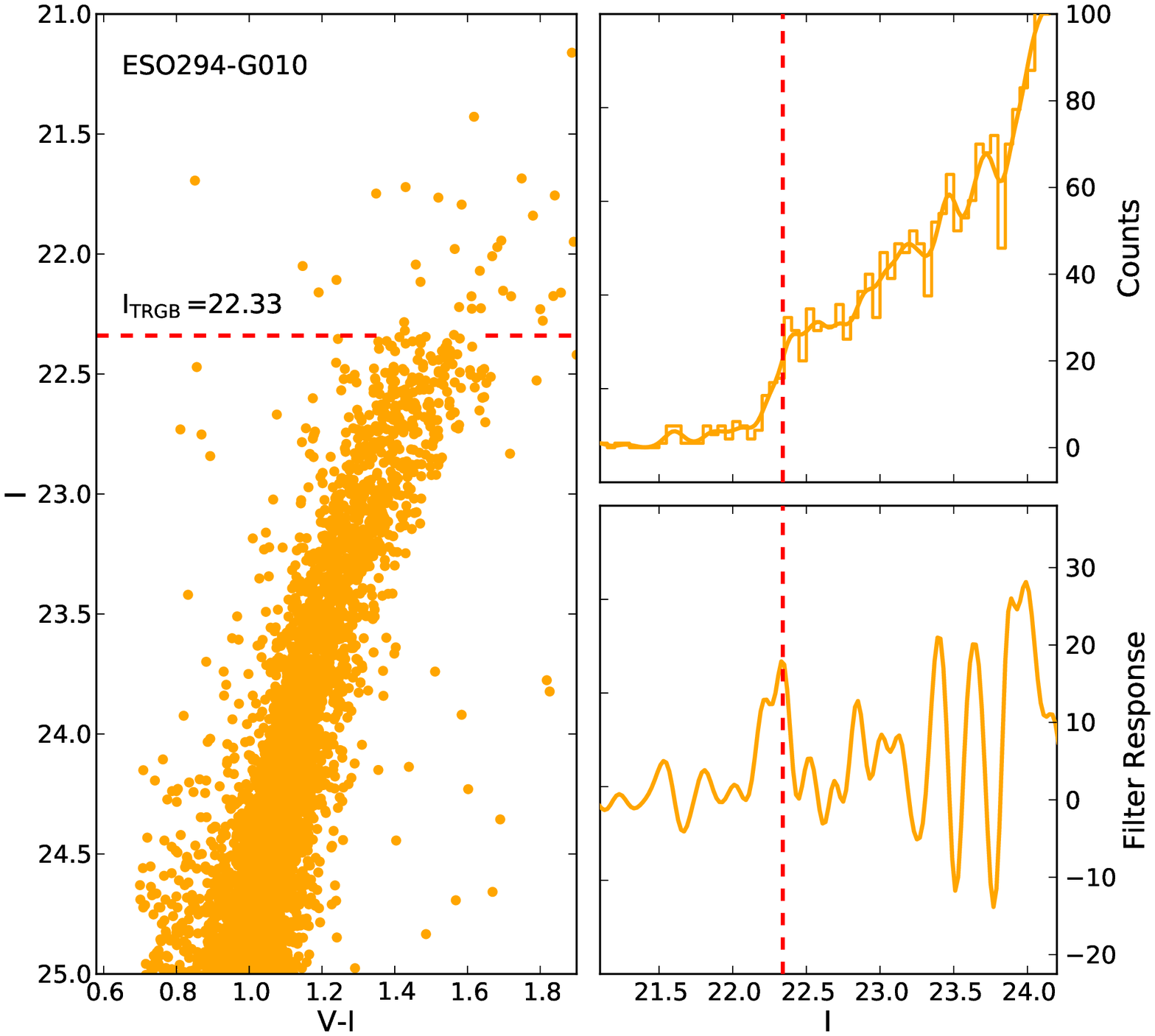}
\caption{Our estimates of TRGB luminosity in the I-band for ESO294-G010. The right-lower panel illustrates the scaled response of a weighted Sobel kernel, [-1, -2, 0, +2, +1] (Madore \& Freedman 1995) on the I-band luminosity function of the RGB stars. \label{fig13}}
\end{figure}

\begin{figure}
\epsscale{1.2}
\plotone{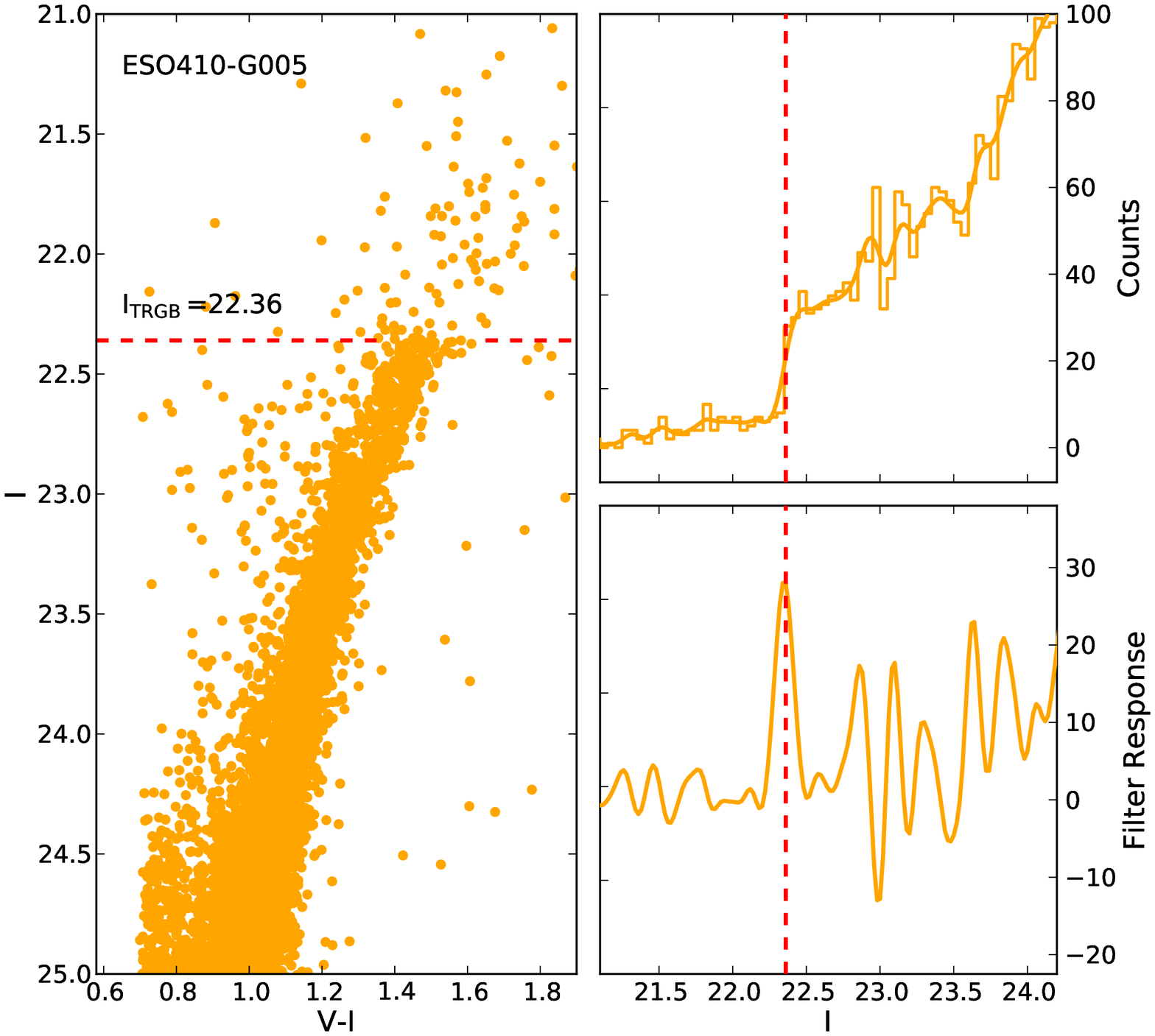}
\caption{Same as Figure 13 but for ESO410-G005.\label{fig14}}
\end{figure}

We have also calculated the distances of the two dTrans galaxies using the I-band magnitude of red giant branch tip (TRGB) in the VI CMDs in order to check the validity of our RR Lyrae distance estimate. We employed a weighted Sobel kernel $[-1, -2, 0, 2, 1]$ (Madore \& Freedman 1995) to detect the TRGB of each galaxy (see Fig 13 \& 14). This yields $I(TRGB) =$ 22.33 $\pm$ 0.05 mag for ESO294-G010 and 22.36 $\pm$ 0.05 for ESO410-G005. Applying the reddening ratio $E(V-I)=$1.38 $E(B-V)$ provided by Tammann, Sandage \& Reindl (2003), and $M^{TRGB}_{I}=$--4.04 $\pm$ 0.12 (Bellazzini et al. 2001, 2004), we obtain $(m-M)_{0} =$ 26.37 $\pm$ 0.13 and 26.38 $\pm$ 0.13 for ESO294-G010 and ESO410-G005 respectively. The errors represent the quadrature sum of the errors in the apparent and absolute I-band TRGB magnitudes. Our distance estimates for ESO294-G010 using two independent methods (RR Lyraes and TRGB) agree very well with each other. While ESO410-G005 shows $\sim$ 0.1 mag discrepancy, this is still consistent within the margin allowed by the errors. The distances to these two galaxies derived in this study show excellent agreement with those measured in previous work such as L13, Da Costa et al. (2010), and  the ``Updated Nearby Galaxy Catalog'' compiled by Karachentsev et al. (2013).

\section{Discusssion}
\subsection{Star Formation History}
The stellar populations of ESO294-G010 and ESO410-G005 share common features in the VI CMDs. However there also exist several noticeable differences in their CMDs indicating that these two dTrans systems might have experienced slightly different star formation histories. The main difference in the CMD features arises from the densities of YMS, BL, and blue HB (BHB) populations, in the sense that the YMS and BL stars in ESO410-G005 appear to be more populous than those in ESO294-G010, while the BHB feature seems relatively stronger in ESO294-G010 than in ESO410-G005. This indicates that ESO294-G010 apparently experienced reduced recent star formation (SF) as compared to ESO410-G005. Our assertion is nicely supported by the recent study of Weisz et al. (2011, W11 hereafter) on the star formation histories (SFHs) of 60 dwarf galaxies in the Local Volume. They used the same ACS/WFC imaging as in this study to extract the SFHs of ESO294-G010 and ESO410-G005 with the synthetic CMD routine developed by Dolphin (2002). W11's results reveal that the two dTrans galaxies exhibit comparable star formation rates (SFRs) during the early epoch (look-back time $>$10 Gyr) of their formation and evolution. Since then, ESO294-G010 has remained mostly dormant and has been sustaining very weak SFRs lower than its lifetime-averaged SFR, while ESO410-G005 shows a sudden enhancement of its SFRs within the most recent $\sim$ 2 Gyr higher than its lifetime-averaged SFR. As a result, ESO294-G010 displays a bluer HB morphology and relatively weak features of luminous YMS and core He burning BL stars compared to ESO410-G005. 

\subsection{Early Chemical Evolution}
According to the standard cosmological models of galaxy formation (White \& Rees 1978; Hernquist \& Quinn 1988, 1989; White \& Frenk 1991), dwarf galaxies have likely formed  through infall of primordial gas into a dark matter halo, and the subsequent star formation episodes determine the paths of their own chemical evolution as typical for galaxies. However the detailed patterns of the chemical evolution of these dwarf galaxies could be different from that of giant galaxies due to their shallower potential wells which lead to more efficient mixing and also make these systems more vulnerable to various kinds of environmental effects such as supernova driven galactic winds, ram pressure stripping, and tidal disruption. If this is the case, what would the early chemical evolution of dwarf galaxies in a low density environment such as the Sculptor group be like? 

To answer this question, we analyze the RR Lyrae MDFs of our target dTrans systems using simple chemical enrichment models (Searle \& Sargent 1972; Pagel \& Patchett 1975; Pagel 1997; Binney \& Merrifield 1998) which depend on the initial heavy element abundance $Z_0$ of the star forming gas, effective nucleosynthetic yield $y$ and loss or gain of star forming gas. The simplest version among the chemical evolution models is the closed-box model with an assumption of no inflow or outflow of gas during the evolution of the system. This model predicts a MDF of a stellar population produced from a star formation episode at a certain age with a gaussian-like exponential function, 

\begin{displaymath}
\frac{\mathit d n}{\mathit d Z} \sim \mathit{e}^{-(Z-Z_{0})/y},   
\frac{\mathit d n}{\mathit d [M/H]} \sim \frac{\mathit Z-Z_{0}}{\mathit y} \mathit{e}^{-(Z-Z_{0})/y}. (3)
\end{displaymath}

In these equations, the yield, $y$ is equivalent to the mean metallicity of the system and $[M/H]$ is the total metal abundance of the system which accounts for the effect of alpha-capture element enhancement [i.e. $ [M/H] = [Fe/H] + log_{10}\left(0.638\times10^{[\alpha/Fe]} + 0.362\right) $; Salaris et al. 1993]. A number of studies (Tolstoy, Hill, \& Tosi 2009 and references therein) have found that the old and metal-poor stars (age $> 10$ Gyr; $[M/H] < -1.6$ dex) in nearby dwarf galaxies exhibit the Galactic halo-like compositions with an enhanced alpha element ratio $\langle [\alpha/Fe] \rangle$ $\sim$ 0.3 dex. Therefore, assuming such an alpha element enhancement should be a reasonable approach. The red solid line in Figure 15 indicates the resulting fits of the simple closed box model with an assumption of a primordial abundance (e.g. $Z_{0}=$0) to our RR Lyrae MDFs of ESO294-G010 and ESO410-G005. We see that the predicted MDFs of the simple closed box model traces the observed MDFs reasonably well; however, in the case of ESO294-G010, the model MDF appears to slightly overestimate the number of metal-poor stars (e.g. $[Fe/H] <$ -2.2 dex). This shortage of the low-Z stars, first recognized as the G-dwarf problem by van den Bergh (1962) and Schmidt(1963), is in fact a common issue for all types of galaxies (Binney \& Merrifield 1998; Harris \& Harris 2000; Pagel \& Patchett 1975; Wheeler et al. 1989). 

\begin{figure}
\epsscale{1.0}
\plotone{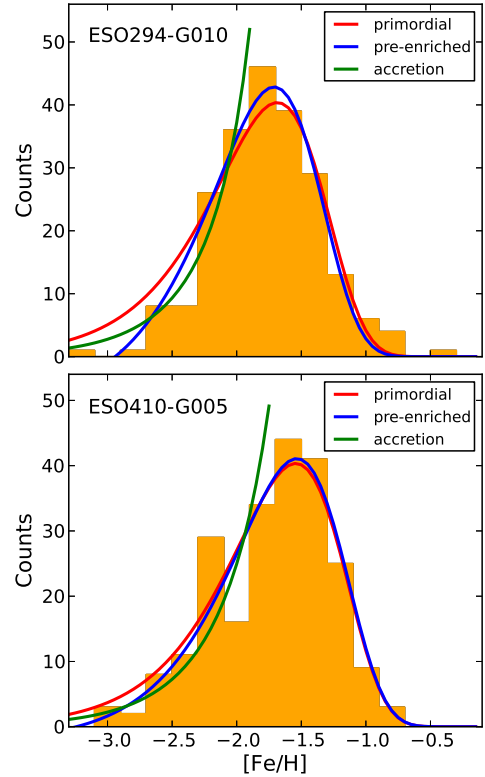}
\caption{Our analysis of the RR Lyrae MDFs with several chemical evolution models. The green, blue, and red curves are the best-fit galactic chemical evolution models to the observed MDFs. \label{fig15}}
\end{figure}

To compensate for this deficit of metal-poor stars, we adopt a pre-enrichment ($Z_{0} \ne 0$) hypothesis for the star forming gas in which the dwarf galaxies were formed. The blue solid lines in Figure 15 illustrate the best-fit pre-enriched closed-box models. For ESO294-G010, a pre-enrichment model with an initial metal abundance of $Z_{0}=0.00004 \pm 4.45E-06$ (i.e. $[M/H]_{0} \sim -2.7$ dex) and a chemical yield of $y=0.00064 \pm 0.00003$ enhances the fit to the metal-poor tail of the observed MDF as compared to the simple closed box model. In the case of ESO410-G005, a pre-enriched system with an initial metal abundance of $Z_{0}=0.00002 \pm 4.43E-6$ (i.e. $[M/H]_{0} \sim -3.0$ dex) and a yield of $y=0.00097 \pm 0.00003$ agrees well with the observed RR Lyrae MDF featuring a comparable quality of fit to the simple closed box model. 

The results of our analysis above naturally lead to the question as to the origin of the pre-enrichment source, and how the conditions in which the pre-enrichment was attained. It has been suggested by Truran \& Cameron (1971) that such pre-enrichment in the star forming gas can be achieved through the process of ``prompt initial enrichment'' or ``initial nucleosynthesis spike'' which includes a preferential pre-galactic or proto-galactic event by high-mass stars during the very early phase of galaxy formation. Indeed our estimates of the initial metal abundances ($Z_{0,ESO294}=0.00004$; $Z_{0,ESO410}=0.00002$) from the best-fit pre-enrichment models are comparable to the initial cosmic metal enrichment levels ($10^{-6} < Z < 10^{-4}$) attained by the end products of population III stars (Schneider et al. 2002). In addition to this, it cannot be ruled out that the proto-galactic bodies of ESO294-G010 and ESO410-G005 could have been enriched by the products from more evolved systems of their nearest neighbor, NGC 55. Both hypotheses seem reasonable to explain the early chemical enrichment histories of our target dTrans galaxies in the low density environment of the Sculptor group.

Another approach to explain the metal-poor tail of the observed MDF is to lift the pure closed-box assumption by introducing an early phase gas inflow (or accretion) from the environment. The simplest form of this ``accretion model'' or ``accreting-box'' modification (Larson 1972; Timmes et al. 1995; Gibson \& Matteucci 1997) assumes that the gas inflow rate is equals to the star formation rate of the system. This means that the mass of star forming gas $M_{g}$ was a constant during the early epoch of galaxy formation. In these configurations, the predicted MDF is simplified to the following hyperbolic function (Binney \& Merrifield 1998; Harris \& Harris 2000; Sarajedini \& Jablonka 2005), 

\begin{displaymath}
\frac{\mathit d n}{\mathit d Z} \sim \frac{\mathit M_g}{y-Z}, 
\frac{\mathit d n}{\mathit d [M/H]} \sim {\mathit M_g} \frac{\mathit Z}{\mathit y-Z}. (4)
\end{displaymath}

In this formulation, during the early phase of evolution the system accumulates the heavy elements very efficiently with a high nucleosynthetic yield. As $Z$ approaches $y$ the dilution effect of the infalling gas starts to catch up to and ultimately squares off the accumulation rate of the heavy elements, thus the system reaches a steady state with no further global chemical enrichment. In Figure 15, we see that the observed metal-poor tails of the RR Lyrae MDFs ($[Fe/H] < -1.7$ dex) of both dwarf galaxies are very well described by our simple accretion models. From the best-fit models, we find $y=0.00126 \pm 0.00003$ and $0.00315 \pm 0.00007$ for ESO294-G010 and ESO410-G005 respectively. As mentioned, this simple accretion model only accounts for the metal-poor stars formed during the very early epoch of the chemical evolution of the system. One of the most probable scenarios to explain the global pattern of the observed MDFs of the two dTrans galaxies is the so called ``two phase scenario'' of chemical evolution (Wyse \& Gilmore 1993; Harris \& Harris 2000; Sarajedini \& Jablonka 2005):  A bulk of stars attributed to the metal-poor tails of the MDFs formed in a very early gas infall phase. Once ESO294-G010 and ESO410-G005 were enriched enough to enter a steady state ($Z \rightarrow y$), both galaxies entered into the closed-box-like phase and the rest of the stars associated with the remaining parts of the MDFs formed in the enriched gas deriving further chemical enrichment. Our analysis of the MDFs of RRL stars (age $>$10 Gyr \& age spread among RR Lyraes in a given stellar system does not exceed $\sim$ 2 Gyr; Lee \& Carney 1999) also suggests that this phase transition should occur within a short period of time.

\subsection{Luminosity-Metallicity Relation}
The luminosity-metallicity (or mass-metallicity) relation of galaxies (L-M relation) is one of the major predictions of chemical evolution models (van den Bergh 1962; Schmidt 1963; Searle \& Sargent 1972; Erb et al. 2006), and it has been confirmed observationally by many studies. The current understanding of the L-M relation based on both theoretical and observational studies suggests that (1) the L-M relation exists among all kinds of galaxies within the Local Volume ($D < 10$ Mpc) regardless of their morphological types (Lequeux et al. 1979; Skillman et al. 1989; Zaritsky et al. 1994; Mateo 1998; Grebel et al. 2003, G03 hereafter; Salzer et al. 2005);  (2) the L-M relation also exists among galaxies at intermediate redshift (i.e. $0.4 < z < 1.0$), and it seems that the L-M relation evolves over cosmological time in the sense that the slope at lower redshifts tends to be flatter than that at higher redshifts (i.e. The metallicity refers to the oxygen nebular emission, $12+log(O/H)$ in this case; Savaglio et al. 2005; Kobulnicky \& Koo 2000; Kobulnicky et al. 2003; Kobulnicky \& Kewley 2004; Maier et al. 2004; Shapley et al. 2005). The L-M relation for the old stellar populations of dwarf satellite galaxies in the Local Group and its neighbors is well established by using mean stellar $[Fe/H]$-values which are based on both spectroscopic and photometric measurements for old RGB stars (see Figure 1. of G03). This L-M relation for nearby galaxies becomes not only a very useful analytical tool for understanding the fundamental differences and a possible evolutionary link between satellite galaxies with different morphological types, but also a basic ladder for the proper interpretation of the L-M relations at high redshifts. 

With the mean metal abundances of the purely ancient stellar populations measured in section 3.3 by using the periods and amplitudes of the RRab stars, we plot ESO294-G010 and ESO410-G005 on the L-M relation of the nearby galaxies in Figure 16. The sample of the known nearby galaxies within a distance of $D \sim 2 $Mpc has been collected from the seminal work of G03 and the latest nearby galaxy catalog of Karachentsev et al. (2013; K13 hereafter). Our compilation (see Table 5) includes the recently discovered ultra-faint dwarf systems (uFds) around the Milky Way, and faint Andromeda dwarf galaxies (Willman et al. 2005a,b; Belokurov et al. 2006; Zucker et al. 2004, 2006a,b, 2007; Irwin et al. 2007,2008; Walsh, Jerjen \& Willman 2007; McConnachie et al. 2008; Martin et al. 2009; Richardson et al. 2011; Bell et al. 2011; Slater et al. 2011), which yields a total of 79 sample dwarf satellite galaxies. We use integrated K-band luminosity ($L_{K}$) values (adopted from Table 2 of K13) as the abscissa instead of using absolute V or B band magnitudes because the $L_{K}$ is a better tracer for the total mass of a given galaxy (Drory et al. 2004).

\begin{figure}
\epsscale{1.2}
\plotone{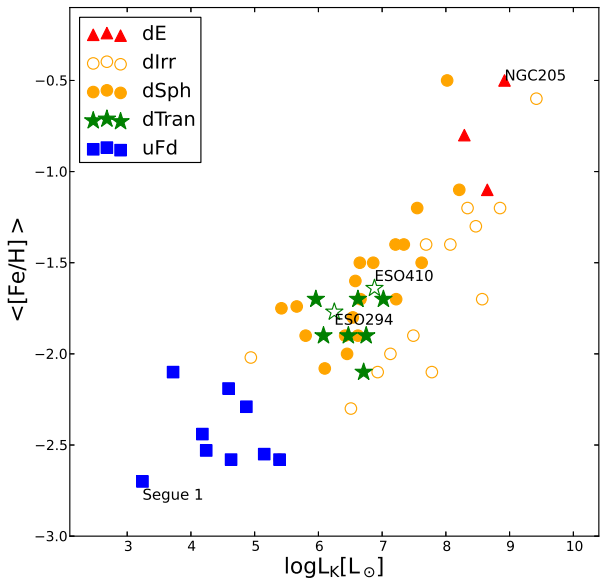}
\caption{Luminosity--Metallicity (L-M) relation for Local Group dwarf galaxies. The data were obtained from the compilation of G03 for the canonical dwarf satellites. For the uFd systems, the data come from the recent work of Norris et al. (2010). ESO294-G010 and ESO410-G005 are shown as open star symbols. The $M_V$--metallicity relation appears to be very well established from the brightest dE, NGC 205, to the faintest MW satellite, Segue 1. Both dTrans in the Sculptor Group trace the average L--M relation of the Local Group dSph galaxies. The typical error in the mean metallicity is about $\sim$ 0.3--0.4 dex.\label{fig16}}
\end{figure}

From Figure 16, we see that all nearby galaxies, from luminous dEs ($red$ $triangle$) to uFds ($blue$ $square$), fall along the typical trend of the L-M relation - more luminous galaxies tend to be more metal-rich.  Noticeable separations between dSphs ($orange$ $dots$)and dIrrs ($orange$ $circles$) are also observed in the sense that the dSphs appear to have higher metal abundances as compared to the dIrrs at the same luminosity level. This implies that at the early epochs of formation (age $> 10$ Gyr), the dSphs should have relatively more active star formation (SFs). As a result, the early chemical enrichment in the dSphs becomes much faster and efficient than in the dIrrs. This scenario of the early SF enhancement in the dSphs is echoed by the work of W11. Using the synthetic CMD analysis (Dolphin 2002), they have drawn the average cumulative SFs for five different morphological types (dSph, dIs, dTrans, dSpirals, and dTidals; see Figure 6 of W11). They have shown that on average the dSphs build up the bulk of their stellar populations (i.e. almost 65 -- 70 $\%$ of their stellar masses) at the very earliest epochs of their formation z $>$ 2 (i.e. age $>$ 10 Gyr), while star formation in the dIrrs remains at the $\sim$ 50 $\%$ level during the same epochs. This fundamental difference between the dSphs and dIrrs in their early star formation and chemical enrichment histories has raised serious doubts about the scenario which involves the morphological transformation dIrrs into dSphs by certain gas removal processes (G03). The locations of our target dTrans galaxies ESO294-G010 and ESO410-G005 in the L-M relation agree with the typical trend of the other dTrans systems ($green$ $star$) such as Phoenix, Tucana, Cetus, LGS 3, KKR 25, DDO 210, and Antlia. Within a luminosity range of 5 $< log (L_{K}/L_{\odot}) <$ 7.5, the L-M relation of the sample dTrans appears to show a closer resemblance to that of the canonical dSphs than dIrrs. In this regard, it has been suggested that a bulk of the dTran galaxies can be considered to be ``$present-day$ $progenitors$'' of the dSph galaxies if the recent star formation of dTrans galaxies were largely suppressed by rapid gas loss (G03).  

\subsection{Galaxy Environments}
The most pivotal differences between dTrans and dSphs galaxies might be the degree of recent star formations (SFs) and the local density environments where these galaxies formed and have been evolving. The morphology-density relation of galaxies in the Local Universe (z $\sim$ 0) has revealed how different types of galaxies tend to be arranged in cluster environments, and provides important clues about how cluster/group environments affects the gas content and global SFR in the galaxy members. In general, gas-deficient early-type galaxies and bulge-dominated lenticular galaxies are preferentially located in the central, densest areas of galaxy clusters. Meanwhile, disk-dominated late-types galaxies (i.e. spirals and irregulars) tend to be more sparsely distributed in the outer regions of these clusters. The morphology-density relation appears to be a ubiquitous phenomenon in cluster environments regardless of their shapes and richness (van der Wel et al. 2010, and references therein). 

Likewise, similar patterns of these morphological segregations have been observed among the dwarf satellite galaxies in the Local Group and other nearby galaxy groups. Figure 17 illustrates the {{\sc H~i}\/} gas content of our sample of dwarf galaxies normalized by their K-band luminosity as a function of the tidal index $\Theta_{5}$. This is basically the same plot as in the 2nd panel of the Fig 16 in K13, but for our 79 volume limited ($D_{\odot} < $2 Mpc) sample dwarf galaxies. We consider this ratio to be a proxy of the gas fraction of the galaxy with respect to its total mass. This is a reasonable approximation not only because the K-band luminosity is an excellent tracer for the total mass of the galaxy but also the $M/L$ values of galaxies in the K-band vary by only a factor of two across a wide range of galaxy morphologies and star formation histories (i.e. the $M/L$ values in the B-band vary up to almost a factor of 10 in the nearby dwarf galaxy samples; Drory et al. 2004 and references therein). 

K13 have provided the formula for the tidal index $\Theta_{5}$ in the following form, 

\begin{displaymath}
\Theta_{5} = log \left( \sum_{n=1}^{5} M_{n}/D^{3}_{in} \right) - 10.96, (5)
\end{displaymath}

where $D_{in}$ is the spatial distance of a neighboring galaxy and $M_{n}$ is its mass. The $\Theta_{5}$ is a measure of the tidal force strength on a given galaxy exerted by the five most influential neighbors. Thus, it can be a robust estimate for the local tidal field strength. In this definition, the high positive $\Theta_{5}$ values indicate a galaxy residing in a dense environment, and the negative values of $\Theta_{5}$ represent isolated galaxies or galaxies in a very low density environment. 

\begin{figure}
\epsscale{1.2}
\plotone{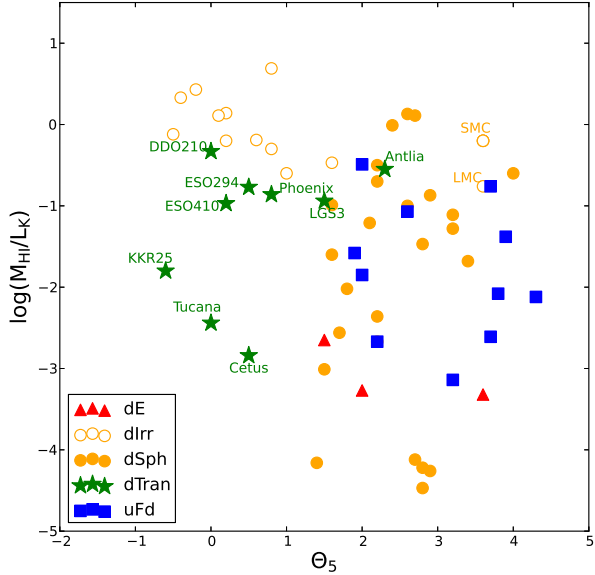}
\caption{The H I gas content (i.e. the H I mass, $M_{H I}$ normalized by K-band luminosity) plotted vs the tidal index, $\Theta_{5}$. The $\Theta_{5}$ is a measure of tidal force strength on a given galaxy exerted by the five most influential neighbors. The high positive $\Theta_{5}$ values indicate a galaxy residing in a dense environment, and the negative values of $\Theta_{5}$ represent isolated galaxies or galaxies in a very low density environment. The majority of gas-rich dIrr galaxies tend to have lower tidal indices except for the LMC/SMC pair, while dSph/uFd/dE galaxies have predominantly high positive values (i.e. There are no early-type dwarf satellites having a $\Theta_{5}$ value less than 1).\label{fig17}}
\end{figure}

In Figure 17, we see a clear morphological segregation by local density environment and gas fraction in the senses like the following: (1) dIrr galaxies are the most gas-abundant systems and there are no dIrrs with a gas fraction $M_{{{\sc H~I}\/}}/L_{K}$ less than 0.1; (2) with the exception of which is known to be one of the closest neighboring galaxy groups the LMC/SMC pair ($\Theta_{5}$$\sim$3.5) and IC 10 ($\Theta_{5}$$\sim$1.6), the majority of the dIrr galaxies are mostly isolated in low density environments ($\Theta_{5} <$ 1). Together with their unusually high 3D velocities ($v_{LMC}=378\pm18$ $km$ $s^{-1}$; Kallivayalil et al. 2006) this could be another piece of evidence supporting the idea that the LMC/SMC pair may not be a canonical satellite system of the Milky Way but an infalling intruder which might be experiencing its first passage around the Milky Way (Besla et al. 2007); and (3) similar to the dIrr galaxies, most dTran galaxies are also found to be isolated but in general they have lower {{\sc H~i}\/} gas fractions as compared to the dIrrs.  

All dSph, uFd and three dE (M32, NGC 205 and NGC 185) galaxies in our sample have relatively high tidal index values $\Theta_{5}>$ 1 indicating that these galaxies have been subjected to relatively higher local tidal fields (i.e. they are located within a galactocentric distance of D$\sim$200 kpc to their giant hosts, Milky Way or M31) than the dTrans and dIrrs galaxies. In terms of the gas content, some of the dSph systems such as Carina, Leo I, Sextans, Draco, Sag dSph, and NGC 147 exhibit the lowest {{\sc H~i}\/} gas fractions [$log(M_{{{\sc H~I}\/}}/L_{K}) < -$4]; however, the rest of the early-type dwarfs are not completely gas-deficient but appear to have a wide range of {{\sc H~i}\/} gas fractions comparable to the levels of the dTran galaxies. Along with the difference in the local density environments, another conspicuous distinction between dSphs and dTrans seems to be the degree of recent SF. According to W11, within the most recent 1 Gyr, the dSph galaxies exhibit largely suppressed SFRs relative to those of dTrans systems. Their estimates of the mean SFHs indicate that the typical dSph and dTran galaxies formed $\sim$ 2\% and 4\% of their total stellar mass respectively within the last 1 Gyr. 

This sudden drop in the recent SFRs of the dSphs might be due to a rapid gas loss which is a quite plausible process in the high local density environments in which these dSphs reside. Similar phenomena have been observed for many larger galaxy group/cluster environments in the Local Universe (z$\sim$0). Recent star formation in galaxies located in denser regions appear to be largely quenched, and these galaxies tend to be more early-types and red (Oemler 1974; Davis \& Geller 1976; Dressler 1980; Postman \& Geller 1984; Wetzel, Tinker \& Conroy 2012). This implies that the effects of environment on galaxy evolution operate in a similar fashion even in the low density environments of the LG and its neighboring groups (i.e. the Sculptor/Maffei groups). 

\section{Summary and Conclusion}
We have presented an extensive analysis of the RR Lyrae stars in two transition type Sculptor group dwarf galaxies, ESO294-G010 and ESO410-G005. Based on the properties of the RR Lyrae stars we present the following results.

\begin{enumerate}
\item We have detected numerous RR Lyrae candidates in both ESO294-G010 ($N_{ab}=$219) and ESO410-G005 ($N_{ab}=$225). Based on the Bailey diagrams, the characteristics of the RR Lyrae stars in these two Sculptor group dTrans appear to follow the typical trend of the RR Lyrae populations of Local Group dwarf satellite galaxies. 

\item We construct the MDFs of the RRab stars using the period-amplitude-[Fe/H] relationship of Alcock et al. (2000). The mean metallicities estimated from the best-fit Gaussians to the MDFs are $<[Fe/H]>=$--1.77 $\pm$ 0.03 (sem) for ESO294-G010, and $<[Fe/H]>=$-1.64 $\pm$ 0.03 (sem) for ESO410-G005 respectively. Our results represent the metallicity values for purely ancient stellar populations in both dTrans galaxies. 

\item The distance of each dTran was calculated using the absolute V magnitudes of the RRab stars. We find $(m-M)_{0}=$ 26.40 $\pm$ 0.07 mag for ESO294-G010 and $(m-M)_{0} =$ 26.33 $\pm$ 0.07 mag for ESO410-G005. We have also calculated the distances of the two dTrans using the I-band magnitudes of TRGBs in order to check the validity of our RR Lyrae distance estimates [$(m-M)_{0,ESO294}=$26.37; $(m-M)_{0,ESO410}=$26.38]. Our distance estimates for ESO294-G010 using two independent methods (RRLs and TRGB) agree very well with each other. In the case of ESO410-G005, there is a $\sim$ 0.1 mag difference but this is consistent  within the margin allowed by the errors.

\item We have compared the RR Lyrae MDFs for ESO294-G010 and ESO410-G005 with several chemical evolution models. For both galaxies, the shapes of the RR Lyrae MDFs are nicely described by pre-enrichment models. This suggests two possible channels for the early chemical evolution of these Sculptor group systems: 1) The ancient stellar populations of our target dwarf galaxies might have formed from the star forming gas which was already enriched through ``prompt initial enrichment'' or an ``initial nucleosynthesis spike'' caused by the very first massive stars, or 2) This pre-enrichment state might have been achieved by the end products from more evolved systems of their nearest neighbor, NGC 55. We also fit a simple accretion model to the RR Lyrae MDFs of each galaxy. The gas infall hypothesis seems to explain the metal-poor tail of the RR Lyrae MDFs indicating that we cannot completely rule out the role of gas accretion on the early chemical evolution of the dTrans.  

\item The L-M relation of our target dTrans ESO294-G010 and ESO410-G005 follow the typical trend of other Local Group dTrans galaxies, such as Phoenix, Tucana, Cetus, LGS 3, KKR 25, DDO 210, and Antlia. Within a luminosity range of 5 $< log (L_{K}/L_{\odot}) <$ 7.5, the L-M relation of the sample dTrans appear to show a closer resemblance to that of the canonical dSphs than dIrrs. This suggests that the bulk of the dTran galaxies can be considered as ``$present-day$ $progenitors$'' of the dSph galaxies if the recent star formation of dTrans galaxies has been largely suppressed by rapid gas loss (G03).  

\item Our examination of the {{\sc H~i}\/} gas content as a function of the tidal index $\Theta_{5}$ for 79 volume limited ($D_{\odot} < $2 Mpc) dwarf galaxies reveals a clear morphological segregation, in the sense that gas-deficient dSphs tend to be located in dense areas while most gas-rich dIrrs (except LMC/SMC pairs) are isolated. Similar to the dIrr galaxies, most dTran galaxies are also found to be isolated but in general they have lower {{\sc H~i}\/} gas fractions as compared to the dIrrs. Our analysis supports the idea that the morphology-density relation appears to be a ubiquitous phenomena in cluster environments regardless of their shapes and richness (van der Wel et al. 2010, and references therein), and the environmental effects on galaxy evolution operate in a similar fashion even in such low density environments of the Local Group or its neighboring groups.
\end{enumerate}

\acknowledgments
We are grateful to the anonymous referee whose comments and suggestions improved the clarity and quality of this paper. 
This work was supported by KASI-Carnegie Fellowship Program jointly managed by Korea Astronomy and Space Science Institute (KASI)
 and the Observatories of the Carnegie Institution for Science.

\clearpage

\clearpage
\begin{deluxetable}{lrrcccccc}
\tabletypesize{\scriptsize}
\tablecaption{Pulsation properties of RR Lyrae stars in ESO294-G010\label{tbl-2}}
\tablewidth{0pt}
\tablehead{
\colhead{ID}            &  \colhead{RA}          &
\colhead{Decl}        &  \colhead{}               & \colhead{$\langle V \rangle$}         &
\colhead{$\langle V \rangle - \langle I \rangle$}   &  \colhead{Period}                &
\colhead{Amp(V)}      &  \colhead{Note}       \\
\colhead{}            &  \colhead{(J2000)}       &
\colhead{(J2000)}     &  \colhead{}              & \colhead{}              &
\colhead{}            &  \colhead{(days)}         &
\colhead{}            &  \colhead{}                           
}
\startdata
 RRab  & WFC1 field &              & &        &       &        &        & \\ 
 31519 & 0:26:27.47 & -41:51:54.93 & & 27.086 & 0.548 & 0.5200 & 0.8116 & \\
 31435 & 0:26:28.37 & -41:51:54.95 & & 26.934 & 0.438 & 0.6760 & 1.0592 &\\
 26874 & 0:26:28.56 & -41:52:12.85 & & 26.759 & 0.394 & 0.6471 & 0.9721 &\\
 29750 & 0:26:29.03 & -41:52:24.77 & & 27.011 & 0.657 & 0.4951 & 1.0102 &\\
 32272 & 0:26:29.42 & -41:50:51.19 & & 27.193 & 0.633 & 0.5932 & 1.0251 &\\
 :           &    :                 &     :                   & &  :            &   :        &    :          &   :          &\\
 \\
 RRc   & WFC1 field &              & &        &       &        &        &\\
 32536 & 0:26:27.76 & -41:52:09.32 & & 26.940 & 0.265 & 0.3004 & 0.6436 &\\
 28014 & 0:26:28.65 & -41:51:03.54 & & 26.792 & 0.546 & 0.3060 & 0.4139 &\\
 29510 & 0:26:31.46 & -41:51:35.01 & & 26.812 & 0.318 & 0.3914 & 0.5809 &\\
 32861 & 0:26:31.94 & -41:52:58.44 & & 27.044 & 0.451 & 0.3699 & 0.4827 &\\
 27866 & 0:26:33.32 & -41:51:34.42 & & 26.729 & 0.434 & 0.3183 & 0.7197 & bl \\
 :           &    :                 &     :                   & &  :            &   :        &    :          &   :          &\\
\\
 RRab  & WFC2 field &              & &        &       &        &        &\\  
 23778 & 0:26:28.85 & -41:49:57.87 & & 26.870 & 0.600 & 0.6373 & 0.4490 &\\
 23713 & 0:26:30.13 & -41:50:17.34 & & 26.823 & 0.484 & 0.5887 & 0.6919 &\\
 25347 & 0:26:31.04 & -41:50:02.74 & & 27.107 & 0.685 & 0.6058 & 1.0376 &\\
 24215 & 0:26:31.34 & -41:49:55.30 & & 26.926 & 0.666 & 0.6529 & 0.7934 &\\
 24335 & 0:26:31.93 & -41:50:41.47 & & 27.007 & 0.429 & 0.5483 & 1.0382 & bl \\
 :           &    :                 &     :                   & &  :            &   :        &    :          &   :          &\\
\\
 RRc   & WFC2 field &              & &        &       &        &        &\\
 23174 & 0:26:33.36 & -41:50:29.40 & & 26.832 & 0.568 & 0.3446 & 0.5975 &\\
 23720 & 0:26:34.14 & -41:50:11.63 & & 26.898 & 0.643 & 0.3781 & 0.4427 &\\
 25657 & 0:26:38.24 & -41:49:48.61 & & 27.108 & 0.605 & 0.4025 & 0.5852 &\\
\enddata
\tablecomments{\footnotesize ``bl'' in the final column indicates $blend$ objects.}
\tablenotetext{**}{The full table will be available in a machine-readable format in the published version}
\end{deluxetable}
\clearpage

\begin{deluxetable}{lrrcccccc}
\tabletypesize{\scriptsize}
\tablecaption{Pulsation properties of RR Lyrae stars in ESO410-G005\label{tbl-3}}
\tablewidth{0pt}
\tablehead{
\colhead{ID}          &  \colhead{RA}          &
\colhead{Decl}        &  \colhead{}            & \colhead{$\langle V \rangle$}         &
\colhead{$\langle V \rangle - \langle I \rangle$}   &  \colhead{Period}      &
\colhead{Amp(V)}      &  \colhead{Note}       \\
\colhead{}            &  \colhead{(J2000)}       &
\colhead{(J2000)}     &  \colhead{}              & \colhead{}              &
\colhead{}            &  \colhead{(days)}        &
\colhead{}            &  \colhead{}                           
}
\startdata
 RRab  & WFC1 field &              & &        &       &        &        &\\ 
 27360 & 0:15:24.34 & -32:10:42.23 & & 26.983 & 0.518 & 0.5624 & 0.9455 &\\
 25183 & 0:15:25.04 & -32:10:40.32 & & 26.830 & 0.636 & 0.6718 & 0.6186 &\\
 25200 & 0:15:25.27 & -32:11:25.31 & & 26.824 & 0.462 & 0.6851 & 0.7559 &\\
 24100 & 0:15:25.34 & -32:11:25.56 & & 26.797 & 0.412 & 0.5933 & 1.1051 &\\
 27060 & 0:15:25.54 & -32:11:31.52 & & 26.926 & 0.412 & 0.5707 & 0.6277 &\\
 :           &    :                 &     :                   & &  :            &   :        &    :          &   :          &\\
\\
 RRc   & WFC1 field &              & &        &       &        &        &\\
 24150 & 0:15:25.57 & -32:11:24.13 & & 26.684 & 0.218 & 0.3242 & 0.5591 &\\
 31623 & 0:15:27.96 & -32:11:25.57 & & 27.143 & 0.394 & 0.2895 & 0.4600 & bl \\
 24390 & 0:15:28.44 & -32:10:58.97 & & 26.759 & 0.574 & 0.2668 & 0.4252 &\\
 26141 & 0:15:28.54 & -32:11:46.15 & & 26.779 & 0.340 & 0.3460 & 0.7136 &\\
 23520 & 0:15:28.66 & -32:11:17.69 & & 26.791 & 0.690 & 0.4048 & 0.6748 &\\
 :           &    :                 &     :                   & &  :            &   :        &    :          &   :          &\\
\\
 RRab  & WFC2 field &              & &        &       &        &        &\\ 
 18627 & 0:15:29.79 & -32:09:45.39 & & 26.987 & 0.622 & 0.5446 & 0.8530 &\\
 15681 & 0:15:30.19 & -32:09:19.01 & & 26.725 & 0.433 & 0.6162 & 0.9297 &\\
 18442 & 0:15:30.40 & -32:10:05.50 & & 26.794 & 0.478 & 0.6037 & 1.2543 &\\
 16370 & 0:15:30.52 & -32:08:38.45 & & 26.845 & 0.617 & 0.6592 & 1.1345 &\\
 20519 & 0:15:30.90 & -32:09:56.61 & & 26.950 & 0.597 & 0.6162 & 1.0357 &\\
 :           &    :                 &     :                   & &  :            &   :        &    :          &   :          &\\
\\
 RRc   & WFC2 field &              & &        &       &        &        &\\
 19124 & 0:15:31.63 & -32:10:15.45 & & 26.809 & 0.479 & 0.3146 & 0.6532 &\\
 16185 & 0:15:31.65 & -32:10:19.04 & & 26.709 & 0.700 & 0.2195 & 0.6250 &\\
 19968 & 0:15:31.99 & -32:10:28.29 & & 26.958 & 0.479 & 0.3597 & 0.7719 & bl \\
 19264 & 0:15:33.55 & -32:09:53.06 & & 26.862 & 0.576 & 0.2885 & 0.5935 &\\
 16866 & 0:15:33.68 & -32:10:26.99 & & 26.783 & 0.667 & 0.3271 & 0.5668 & bl \\
 :           &    :                 &     :                   & &  :            &   :        &    :          &   :          &\\  
\enddata
\tablecomments{\footnotesize ``bl'' in the final column indicates $blend$ objects.}
\tablenotetext{**}{The full table will be available in a machine-readable format in the published version}
\end{deluxetable}
\clearpage

\begin{deluxetable}{lrrcccccc}
\tabletypesize{\scriptsize}
\tablecaption{Pulsation properties of Luminous Variables\label{tbl-4}}
\tablewidth{0pt}
\tablehead{
\colhead{ID}          &  \colhead{RA}          &
\colhead{Decl}        &  \colhead{}            & \colhead{$\langle V \rangle$}         &
\colhead{$\langle V \rangle - \langle I \rangle$}   &  \colhead{Period}      &
\colhead{Amp(V)}      &  \colhead{Note}       \\
\colhead{}            &  \colhead{(J2000)}       &
\colhead{(J2000)}     &  \colhead{}              & \colhead{}              &
\colhead{}            &  \colhead{(days)}        &
\colhead{}            &  \colhead{}                           
}
\startdata
 ESO294-G010 & WFC1 field &              & &        &       &        &        &\\ 
  3817 & 0:26:32.12 & -41:51:49.45 & & 25.377 & 0.543 & 1.8274 & 1.0387 & bl \\
 14561 & 0:26:32.44 & -41:52:24.09 & & 26.038 & 0.393 & 0.4650 & 0.7667 &  \\
 23638 & 0:26:32.62 & -41:52:22.60 & & 26.482 & 0.282 & 0.4418 & 0.5510 &  \\
 21370 & 0:26:33.31 & -41:51:35.21 & & 26.521 & 0.561 & 0.5339 & 0.4820 & bl \\
  1660 & 0:26:33.52 & -41:51:34.13 & & 24.784 & 0.501 & 0.8735 & 0.3924 & bl \\
  5048 & 0:26:33.61 & -41:51:47.51 & & 25.492 & 0.483 & 1.4140 & 0.6971 & bl \\
 18994 & 0:26:33.68 & -41:51:20.84 & & 26.278 & 0.250 & 0.4283 & 0.5306 & bl \\
 19596 & 0:26:33.79 & -41:51:22.39 & & 26.378 & 0.480 & 0.5473 & 0.2845 & bl \\
 20461 & 0:26:33.80 & -41:51:42.26 & & 26.302 & 0.417 & 0.5473 & 0.7387 & bl \\
  1226 & 0:26:33.94 & -41:51:13.44 & & 24.503 & 0.511 & 1.9431 & 0.5900 & bl \\
  5366 & 0:26:34.24 & -41:51:19.52 & & 25.588 & 0.593 & 1.3464 & 0.7185 & bl \\
 15933 & 0:26:34.36 & -41:51:51.62 & & 26.204 & 0.416 & 0.5698 & 0.5743 & bl \\
 ESO294-G010 & WFC2 field &              & &        &       &        &        &\\ 
  2754 & 0:26:32.62 & -41:50:19.13 & & 25.287 & 0.649 & 1.5201 & 0.6490 &  \\
  2412 & 0:26:33.07 & -41:50:47.32 & & 24.933 & 0.390 & 1.1971 & 0.5833 & bl \\
  3249 & 0:26:33.21 & -41:50:55.11 & & 25.395 & 0.594 & 0.4683 & 0.2520 & bl \\
 10896 & 0:26:33.31 & -41:50:53.40 & & 26.136 & 0.482 & 0.7133 & 0.8989 & bl \\
  2432 & 0:26:33.69 & -41:50:59.90 & & 24.967 & 0.473 & 1.2844 & 1.1950 & bl \\
 14201 & 0:26:34.19 & -41:51:00.50 & & 26.209 & 0.485 & 0.6302 & 0.3131 & bl\\
  3942 & 0:26:34.74 & -41:50:10.73 & & 25.642 & 0.631 & 1.2321 & 0.7236 &  \\
\\
 ESO410-G005 & WFC1 field &              & &        &       &        &        &\\ 
  6202 & 0:15:28.56 & -32:11:31.48 & & 25.707 & 0.540 & 0.9203 & 0.6795 &  \\
  1756 & 0:15:29.54 & -32:11:03.79 & & 24.746 & 0.547 & 1.5094 & 1.2808 & bl \\
  2067 & 0:15:29.97 & -32:11:11.94 & & 24.902 & 0.523 & 0.9324 & 0.4800 & bl \\
  1409 & 0:15:30.29 & -32:11:13.20 & & 24.417 & 0.390 & 1.1129 & 0.4249 & bl \\
  2387 & 0:15:30.50 & -32:10:52.73 & & 25.025 & 0.523 & 0.7383 & 0.5900 & bl \\
   973 & 0:15:30.91 & -32:10:55.99 & & 24.268 & 0.419 & 1.7204 & 1.1464 & bl \\
  1959 & 0:15:31.10 & -32:11:03.09 & & 24.887 & 0.652 & 1.7143 & 0.5899 & bl \\
  2278 & 0:15:31.49 & -32:10:53.14 & & 25.039 & 0.538 & 1.2008 & 0.6843 & bl \\
  1147 & 0:15:31.85 & -32:10:52.33 & & 24.207 & 0.441 & 1.1759 & 0.6793 & bl \\
  2803 & 0:15:32.05 & -32:10:51.68 & & 25.159 & 0.630 & 1.2684 & 0.9685 & bl \\
  9037 & 0:15:32.71 & -32:11:22.99 & & 25.987 & 0.606 & 0.5413 & 0.5250 &  \\
  1553 & 0:15:32.76 & -32:10:57.75 & & 24.590 & 0.447 & 1.0492 & 0.3652 & bl \\
ESO410-G005 & WFC2 field &              & &        &       &        &        &\\ 
  1413 & 0:15:32.33 & -32:10:31.04 & & 25.022 & 0.418 & 0.8278 & 0.4415 & bl \\
  1023 & 0:15:32.51 & -32:10:18.59 & & 24.694 & 0.412 & 0.7997 & 0.5748 & bl \\
  1105 & 0:15:32.78 & -32:10:34.68 & & 24.840 & 0.570 & 1.0211 & 0.5039 & bl \\
  1546 & 0:15:33.46 & -32:10:51.05 & & 25.130 & 0.477 & 1.0981 & 0.4714 & bl \\
  1923 & 0:15:33.61 & -32:10:42.03 & & 25.261 & 0.484 & 0.7039 & 0.3242 & bl \\
  6160 & 0:15:33.63 & -32:09:17.19 & & 26.044 & 0.486 & 0.6571 & 0.4327 & bl \\
   879 & 0:15:33.99 & -32:10:22.82 & & 24.603 & 0.476 & 1.0377 & 0.8358 &  \\ 
\enddata
\tablecomments{\footnotesize ``bl'' in the final column indicates $blend$ objects.}
%\tablenotetext{**}{The full table will be available in a machine-readable format in the published version}
\end{deluxetable}
\clearpage

\LongTables
\begin{deluxetable}{llllrccllrc}
\tabletypesize{\scriptsize}
\tablecaption{Physical Properties of Nearby Dwarf Satellite Galaxies \label{tbl-5}}
\tablewidth{0pt}
\tablehead{
\colhead{Object}      & \colhead{RA (J2000)}    & \colhead{Dec. (J2000)}              & 
\colhead{$D_{\odot}$} & \colhead{$M_{B}$}       & \colhead{$\langle [Fe/H] \rangle$}  &
\colhead{Ref}         & \colhead{$log M_{H I}$} & \colhead{$log L_{K}$}               &
\colhead{$\Theta_{5}$}& \colhead{Type}
}
\startdata
MW Group     &              &           &       &       &                   &      &        &        &      &          \\
SMC          &  00 52 38.0  & -72 48 01 &  0.06 & -16.5 &  -1.20 $\pm$ 0.40 &   1  &  8.65  &  8.85  & 3.6  &  dIrr    \\
Sculptor     &  01 00 09.4  & -33 42 33 &  0.09 &  -9.8 &  -1.50 $\pm$ 0.50 &   1  &  5.39  &  6.86  & 2.8  &  dSph    \\
Phoenix      &  01 51 06.3  & -44 26 41 &  0.44 &  -9.6 &  -1.90 $\pm$ 0.40 &   1  &  5.22  &  6.08  & 0.8  &  dTran   \\
Segue 2      &  02 19 16.0  & +20 10 31 &  0.03 &  -2.3 &     -             &   -  &    -   &  3.88  & 3.8  &  uFd     \\
Fornax       &  02 39 54.7  & -34 31 33 &  0.14 & -11.5 &  -1.20 $\pm$ 0.50 &   1  &  5.19  &  7.55  & 2.2  &  dSph    \\
LMC          &  05 23 34.6  & -69 45 22 &  0.05 & -17.9 &  -0.60 $\pm$ 0.50 &   1  &  8.66  &  9.42  & 3.6  &  dIrr    \\
Carina       &  06 41 36.7  & -50 57 58 &  0.10 & -9.0  &  -1.80 $\pm$ 0.30 &   1  &  2.32  &  6.54  & 2.8  &  dSph    \\
UMa II       &  08 51 30.0  & +63 07 48 &  0.03 & -3.1  &  -2.44 $\pm$ 0.53 &   2  &  2.80  &  4.18  & 3.9  &  uFd     \\
LeoT         &  09 34 53.4  & +17 03 05 &  0.42 & -6.7  &  -2.02 $\pm$ 0.54 &   4  &  5.63  &  4.94  & 0.8  &  dIrr    \\  
SexB         &  10 00 00.1  & +05 19 56 &  1.36 & -14.0 &  -2.10 $\pm$ 0.40 &   1  &  7.66  &  7.78  & -0.5 &  dIrr    \\  
Segue 1      &  10 07 03.2  & +16 04 25 &  0.02 & -0.7  &  -2.70 $\pm$ 0.70 &   2  &  1.12  &  3.24  & 4.3  &  uFd     \\
LeoI         &  10 08 26.9  & +12 18 29 &  0.25 & -11.0 &  -1.40 $\pm$ 0.50 &   1  &  3.18  &  7.34  & 1.4  &  dSph    \\ 
SexA         &  10 11 00.8  & -04 41 34 &  1.32 & -13.9 &  -1.90 $\pm$ 0.40 &   1  &  7.82  &  7.49  &-0.4  &  dIrr    \\ 
Sextans      &  10 13 03.0  & -01 36 52 &  0.09 &  -8.7 &  -1.90 $\pm$ 0.40 &   1  &  2.30  &  6.42  & 2.7  &  dSph    \\
UMa I        &  10 34 52.8  & +51 55 12 &  0.10 &  -4.8 &  -2.29 $\pm$ 0.50 &   2  &  3.80  &  4.87  & 2.6  &  uFd     \\ 
Willman1     &  10 49 21.0  & +51 03 00 &  0.04 &  -1.9 &  -2.10            &   5  &  2.96  &  3.72  & 3.7  &  uFd     \\ 
LeoII        &  11 13 29.2  & +22 09 17 &  0.21 &  -9.1 &  -1.60 $\pm$ 0.50 &   1  &  4.02  &  6.58  & 1.7  &  dSph    \\ 
LeoV         &  11 31 09.6  & +02 13 12 &  0.18 &  -3.8 &       -           &  -   &  2.88  &  4.46  & 1.9  &  uFd     \\
LeoIV        &  11 32 57.0  & -00 32 00 &  0.16 &  -4.2 &  -2.58 $\pm$ 0.72 &   2  &  2.78  &  4.63  & 2.0  &  uFd     \\ 
ComaI        &  12 26 59.0  & +23 54 15 &  0.04 &  -3.2 &  -2.53 $\pm$ 0.40 &   2  &  1.63  &  4.24  & 3.7  &  uFd     \\
CVnII        &  12 57 10.0  & +34 19 15 &  0.16 &  -4.1 &  -2.19 $\pm$ 0.54 &   2  &  4.10  &  4.59  & 2.0  &  uFd     \\ 
CVnI         &  13 28 03.5  & +33 33 21 &  0.22 & -7.9  & -2.08 $\pm$ 0.41  &   2  &  4.50  &  6.10  & 1.6  &  dSph    \\  
BootesIII    &  13 57 07.4  & +26 46 30 &  0.05 & -5.9  &      -            &  -   &   -    &  5.29  & 3.6  &  uFd     \\
BootesII     &  13 58 00.0  & +12 50 00 &  0.04 & -1.9  &      -            &  -   & 1.62   &  3.70  & 3.8  &  uFd     \\ 
BootesI      &  14 00 00.0  & +14 30 00 &  0.07 & -5.5  & -2.55 $\pm$ 0.37  &   2  &  2.01  &  5.15  & 3.2  &  uFd     \\
UMin         &  15 09 11.3  & +67 12 52 &  0.06 & -7.1  & -1.90 $\pm$ 0.70  &   2  &  4.52  &  5.80  & 3.2  &  dSph    \\ 
Hercules     &  16 31 02.0  & +12 47 30 &  0.15 & -6.1  & -2.58 $\pm$ 0.47  &   2  &  2.72  &  5.39  & 2.2  &  uFd     \\
Draco        &  17 20 01.4  & +57 54 34 &  0.08 & -8.7  & -2.00 $\pm$ 0.70  &   1  &  2.19  &  6.45  & 2.9  &  dSph    \\
Sag dSph     &  18 55 03.1  & -30 28 42 &  0.02 & -12.7 & -0.50 $\pm$ 0.80  &   1  &  1.99  &  8.02  & 5.3  &  dSph    \\ 
Sag dIrr     &  19 29 59.0  & -17 40 41 &  1.04 & -11.5 & -2.30 $\pm$ 0.40  &   1  &  6.94  &  6.51  & -0.2 &  dIrr    \\ 
NGC6822      &  19 44 57.7  & -14 48 11 &  0.50 & -15.2 & -1.20 $\pm$ 0.40  &   1  &  8.15  &  8.34  & 0.6  &  dIrr    \\  
Tucana       &  22 41 49.0  & -64 25 12 &  0.88 & -9.2  & -1.70 $\pm$ 0.20  &   1  &  4.18  &  6.62  & 0.0  &  dTran   \\ 
PiscesII     &  22 58 31.0  & +05 57 09 &  0.18 & -4.4  &       -           &  -   &  -     &  4.70  & 1.9  &  uFd     \\ 
\\
M31 Group    &              &           &       &       &                   &      &        &        &      &          \\
WLM          &  00 01 58.1  & -15 27 40 & 0.97  & -14.1 &  -1.40 $\pm$ 0.40 &    1 &  7.83  &  7.69  &  0.2 &  dIrr    \\ 
And XVIII    &  00 02 14.5  & +45 05 20 & 1.36  &  -9.1 &     -             &  -   &   -    &  6.60  &  0.5 &  dSph    \\
And XX       &  00 07 30.7  & +35 07 56 & 0.80  &  -5.8 &    -              &  -   &  -     &  5.26  &  2.4 &  dSph    \\
And XIX      &  00 19 32.1  & +35 02 37 & 0.93  &  -8.3 &     -             &  -   &  -     &  6.28  &  1.9 &  dSph    \\
IC10         &  00 20 24.5  & +59 17 30 & 0.66  & -16.0 &  -1.30 $\pm$ 0.40 &   1  &  8.00  &  8.47  &  1.6 &  dIrr    \\
And XXVI     &  00 23 45.6  & +47 54 58 & 0.76  &  -6.5 &     -             & -    &  -     &  5.54  &  2.9 &  dSph    \\  
Cetus        &  00 26 11.0  & -11 02 40 & 0.78  & -10.2 &  -1.70 $\pm$ 0.20 &   1  &  4.18  &  7.02  &  0.5 &  dTran   \\ 
And XXV      &  00 30 08.9  & +46 51 07 & 0.81  & -9.1  &      -            &  -   &  -     &  6.58  &  2.9 &  dSph    \\ 
NGC147       &  00 33 11.6  & +48 30 28 & 0.76  & -14.8 &  -1.10 $\pm$ 0.40 &  1   &  3.74  &  8.21  &  2.8 &  dSph    \\ 
And III      &  00 35 33.8  & +36 29 52 & 0.75  &  -9.3 &  -1.70 $\pm$ 0.20 &   1  &  5.55  &  6.66  &  3.2 &  dSph    \\ 
And XXX      &  00 36 34.9  & +49 38 48 & 0.68  &   -   &    -              &   -  &  -     &   -    &  2.4 &   -      \\
And XVII     &  00 37 07.0  & +44 19 20 & 0.74  & -7.0  &     -             &  -   &  5.54  &  5.74  &  3.6 &  dSph    \\
And XXVII    &  00 37 27.1  & +45 23 13 & 0.83  & -7.3  &     -             &  -   & -      &  5.88  &  3.1 &  dSph    \\  
NGC185       &  00 38 58.0  & +48 20 10 & 0.61  & -14.7 &  -0.80 $\pm$ 0.40 &   1  &  5.02  & 8.29   &  2.0 &  dE      \\  
NGC205       &  00 40 22.5  & +41 41 11 & 0.82  & -16.1 &  -0.50 $\pm$ 0.50 &   1  &  5.60  &  8.92  &  3.6 &  dE      \\ 
M32          &  00 42 42.1  & +40 51 59 & 0.49  & -14.8 &  -1.10 $\pm$ 0.60 &    1 &  6.00  &  8.65  &  1.5 &  dE      \\ 
And I        &  00 45 40.0  & +38 02 14 & 0.73  & -10.7 &  -1.40 $\pm$ 0.20 &   1  &  5.53  &  7.21  &  3.4 &  dSph    \\  
And XI       &  00 46 20.0  & +33 48 05 & 0.73  &  -6.2 &  -1.75 $\pm$ 0.18 &   3  &  5.53  &  5.42  &  2.7 &  dSph    \\ 
And XII      &  00 47 27.0  & +34 22 29 & 0.83  &  -6.4 &     -             &  -   &  5.64  &  5.51  &  2.6 &  dSph    \\ 
And XIV      &  00 51 35.0  & +29 41 49 & 0.73  &  -7.7 &      -            &  -   &  5.53  &  6.03  &  2.2 &  dSph    \\
And XIII     &  00 51 51.0  & +33 00 16 & 0.84  &  -6.8 &  -1.74 $\pm$ 0.18 &  3   &  5.65  &  5.66  &  2.4 &  dSph    \\ 
And IX       &  00 52 52.8  & +43 12 00 & 0.79  &  -8.1 &       -           &  -   &  5.60  &  6.20  &  4.0 &  dSph    \\ 
And XVI      &  00 59 29.8  & +32 22 36 & 0.52  &  -8.2 &      -            & -    &  5.24  &  6.23  &  1.6 &  dSph    \\ 
LGS 3        &  01 03 55.0  & +21 53 06 & 0.65  &  -9.3 &  -1.70 $\pm$ 0.30 & 1    &  5.02  &  5.96  &  1.5 &  dTran   \\ 
IC1613       &  01 04 47.8  & +02 08 00 & 0.73  & -14.5 &  -1.40 $\pm$ 0.30 & 1    &  7.77  &  8.07  &  0.8 &  dIrr    \\ 
And X        &  01 06 33.7  & +44 48 16 & 0.63  &  -7.9 &      -            & -    &  5.40  &  6.10  &  2.2 &  dSph    \\ 
And V        &  01 10 17.1  & +47 37 41 & 0.81  &  -9.2 &  -1.90 $\pm$ 0.10 & 1    &  5.62  &  6.62  &  2.6 &  dSph    \\ 
And XV       &  01 14 18.7  & +38 07 03 & 0.76  &  -8.7 &       -           &  -   &  5.56  &  6.43  &  2.9 &  dSph    \\ 
And II       &  01 16 29.8  & +33 25 09 & 0.65  &  -9.2 &  -1.50 $\pm$ 0.30 & 1    &  5.44  &  6.65  &  2.1 &  dSph    \\ 
And XXIV     &  01 18 30.0  & +46 21 58 & 0.60  &  -7.0 &       -           &  -   &   -    &  5.77  &  1.9 &  dSph    \\ 
And XXII     &  01 27 40.0  & +28 05 25 & 0.79  &  -6.0 &       -           &  -   & -      &  5.36  &  2.3 &  dSph    \\  
And XXIII    &  01 29 21.8  & +38 43 08 & 0.73  &  -9.5 &       -           &   -  &   -    &  6.75  &  2.5 &  dSph    \\ 
LeoA         &  09 59 26.4  & +30 44 47 & 0.81  & -11.7 &  -2.10 $\pm$ 0.40 &   1  &  7.04  &  6.93  &  0.1 &  dIrr    \\  
KKR25        &  16 13 47.6  & +54 22 16 & 1.86  & -9.4  &  -2.10 $\pm$ 0.40 &   1  &  4.91  &  6.71  & -0.6 &  dTran   \\ 
DDO210       &  20 46 51.8  & -12 50 53 & 0.94  & -11.1 &  -1.90 $\pm$ 0.30 &   1  &  6.42  &  6.75  &  0.0 &  dTran   \\ 
IC5152       &  22 02 41.9  & -51 17 43 & 1.97  & -15.6 &  -1.40 $\pm$ 0.50 &    1 &  8.02  &  8.72  & -0.7 &  BCD     \\ 
And XXVIII   &  22 32 41.2  & +31 12 58 & 0.65  &  -7.7 &      -            & -    &   -    &  6.04  &  1.2 &  dTran   \\ 
Cas dSph     &  23 26 31.8  & +50 40 32 & 0.79  & -11.7 &  -1.50 $\pm$ 0.20 &    1 &  5.60  &  7.62  &  1.8 &  dSph    \\ 
Pegasus      &  23 28 34.1  & +14 44 48 & 0.76  & -11.5 &  -2.00 $\pm$ 0.30 &    1 &  6.53  &  7.13  &  1.0 &  dIrr    \\ 
Peg dSph     &  23 51 46.4  & +24 35 10 & 0.82  & -10.7 &  -1.70 $\pm$ 0.20 &   1  &  4.21  &  7.22  &  1.5 &  dSph    \\ 
And XXI      &  23 54 47.7  & +42 28 15 & 0.86  &  -9.3 &      -            &   -  &   -    &  6.66  &  2.2 &  dSph    \\ 
And XXIX     &  23 58 55.6  & +30 45 20 & 0.73  &  -7.5 &     -             &   -  & -      &  5.96  &  2.0 &  dSph    \\ 
\\
Antilia Group&              &           &       &       &                   &      &        &        &      &          \\
NGC3109      &  10 03 07.2  & -26 09 36 & 1.32  & -15.7 &  -1.70 $\pm$ 0.40 &   1  &   8.37 &  8.57  &  0.2 &  dIrr    \\
Antlia       &  10 04 04.0  & -27 19 55 & 1.32  &  -9.8 &  -1.90 $\pm$ 0.20 &   1  &   5.92 &  6.47  &  2.3 &  dTran   \\  
\\
Sculptor Group &            &           &       &       &                   &      &        &        &       &          \\
ESO410-G005  & 00 15 31.4   & -32 10 48 & 1.82  & -11.6 &  -1.64 $\pm$ 0.03 & this study  &  5.91 &  6.88 & 0.2 & dTran\\
ESO294-G010  & 00 26 33.3   & -41 51 20 & 1.88  & -10.9 &  -1.77 $\pm$ 0.03 & this study  &  5.48 &  6.25 & 0.5 & dTran\\ 
\enddata
%\tablecomments{\footnotesize Ref indicates the reference for the quoted metallicity values: 1. Grebel, E. K., Gallagher, J. S., III, \& Harbeck, D. 2003; 2. Norris, J. E. et al. 2010, ApJ, 723, 1632; 3.  Yang \& Sarajedini 2012 (YS12); 4. Kirby, E. N., Simon, J. D., Geha, M., Guhathakurta, P., \& Frebel, A. 2008; 5. Willman, B. et al. 2011}
\tablerefs{\footnotesize 1. Grebel, E. K., Gallagher, J. S., III, \& Harbeck, D. 2003; 2. Norris, J. E. et al. 2010, ApJ, 723, 1632; 3.  Yang \& Sarajedini 2012 (YS12); 4. Kirby, E. N., Simon, J. D., Geha, M., Guhathakurta, P., \& Frebel, A. 2008; 5. Willman, B. et al. 2011}
\end{deluxetable}


\begin{thebibliography}{}
\bibitem[Alcock et al.(2000)]{alc00} Alcock, C., et al. 2000, AJ, 119, 2194
\bibitem[Binney \& Merrifield (1998)]{bm98} Binney, J., \& Merrifield, M. 1998, Galactic Astronomy (Princeton, NJ : Princeton University Press) 
\bibitem[Bell et al.(2011)]{bsm11} Bell, E. F., Slater, C. T., \& Martin, N. F. 2011, ApJL, 742, 15
\bibitem[Bellazzini et al.(2001)]{bfp01} Bellazzini, M., Ferraro, F. R., \& Pancino, E. 2001, ApJ, 556, 635
\bibitem[Bellazzini et al.(2004)]{bfs04} Bellazzini, M., Ferraro, F. R., Sollima, A., Pancino, E., \& Origlia, L. 2004, A\&A, 424, 199
\bibitem[Belokurov et al.(2006)]{bel06} Belokurov, V. et al. 2006, ApJL, 647, 111
%\bibitem[Belokurov et al.(2007)]{bel07} Belokurov, et al. 2007
\bibitem[Bernard et al.(2008)]{brd08} Bernard, E. J. et al. 2008, ApJ, 678L, 21
\bibitem[Bernard et al.(2009)]{brd09} Bernard, E. J. et al. 2009, ApJ, 699, 1742
\bibitem[Belsa et al.(2007)]{bes07} Besla, G., Kallivayalil, N., Hernquist, L., Robertson, B., Cox, T. J., van der Marel, R. P., \& Alcock, C. 2007, ApJ
, 668, 949
\bibitem[Bouchard et al.(2005)]{bdj05} Bouchard, A., Da Costa, G. S., Jerjen, H, \& Ott, J. 2005, in IAUC. 198, Near-Field Cosmology with Dwarf Elliptical 
Galaxies (Les Diablerets, Switzerland), 255
\bibitem[Cacciari et al.(2005)]{ccc05} Cacciari, C., Corwin, T. M., \& Carney, B. W. 2005, AJ, 129, 267
\bibitem[Castellani \& Quarta (1987)]{cq87} Castellani, V. \& Quarta, M. L. 1987, A\&AS, 71, 1
\bibitem[Charboyer (1999)]{cha99} Chaboyer, B. 1999, in Heck A., Caputo F., eds, Astrophysics and Space Science Library Vol. 237, Post-Hipparcos Cosmi
c Candles. Kluwer, Dordrecht, p.111
\bibitem[Clement \& Rowe (2000)]{cr00} Clement, C. M., \& Rowe, J. 2000, AJ, 120, 2579
\bibitem[Da Costa et al.(2010)]{drj10} Da Costa, G. S., Rejkuba, M., Jerjen, H., \& Grebel, E. K. 2010, ApJL, 708, 121
\bibitem[Davis \& Geller (1976)]{dg76} Davis, M., \& Geller, M. J. 1976, ApJ, 208, 13
\bibitem[de Vaucouleurs et al.(1991)]{ddc91} de Vaucouleurs, G., de Vaucouleurs, A., Corwin, H. G., Jr., Buta, R. J., Paturel, G., \& Fouqu�, P. 1991, Third Refe
rence Catalogue of Bright Galaxies. Volume I: Explanations and references. Volume II: Data for galaxies between $0^h$ and $12^h$.
 Volume III: Data for galaxies between $12^h$ and $24^h$. (New York, NY (USA) : Springer), ISBN 0-387-97552-7
\bibitem[Dolphin (2000)]{dol00} Dolphin, A. E. 2000, PASP, 112, 1383
\bibitem[Dolphin (2002)]{dol02} Dolphin, A. E. 2002, MNRAS, 332, 91
\bibitem[Dolphin et al.(2004)]{dol04} Dolphin, A. E. et al. 2004, AJ, 127, 875
\bibitem[Dotter et al.(2008)]{dcj08} Dotter, A., Chaboyer, B., Jevremovi?, D., Kostov, V., Baron, E., \& Ferguson, J. W. 2008, ApJS, 178, 89
\bibitem[Dressler (1980)]{dre80} Dressler, A. 1980, ApJ, 236, 351
\bibitem[Drory et al.(2004)]{dbf04} Drory, N., Bender, R., Feulner, G., Hopp, U., Maraston, C., Snigula, J., \& Hill, G. J. 2004, ApJ, 608, 742
\bibitem[Erb et al.(2006)]{esp06} Erb, D. K., Shapley, A. E., Pettini, M., Steidel, C. C., Reddy, N. A., \& Adelberger, K. L. 2006, ApJ, 644, 813
%\bibitem[Garnett et al.(1997)]{gar97} Garnett, et al. 1997
\bibitem[Gallart et al.(2004)]{gal04} Gallart, C. et al. 2004, ApJ, 127, 1486
\bibitem[Gibson \& Matteucci (1997)]{gm97} Gibson, B. K., \& Matteucci, F. 1997, MNRAS, 291, 8
\bibitem[Grebel et al.(2003)]{ggh03} Grebel, E. K., Gallagher, J. S., III, \& Harbeck, D. 2003, AJ, 125, 1926
\bibitem[Guldenschuh et al.(2005)]{gul05} Guldenschuh, K. A., et al. 2005, PASP, 117, 721
\bibitem[Harris \& Harris (2000)]{hh00} Harris, G. L. H., \& Harris, W. E. 2000, AJ, 120, 2423
\bibitem[Hernquist \& Quinn (1988)]{hq88} Hernquist, L., \& Quinn, P. J. 1988, ApJ, 331, 682
\bibitem[Hernquist \& Quinn (1989)]{hq89} Henrquist, L., \& Quinn, P. J. 1989, ApJ, 342, 1
\bibitem[Irwin et al.(2007)]{irw07} Irwin, M. J. et al. 2007, ApJL, 656, 13
\bibitem[Irwin et al.(2008)]{ifh08} Irwin, M. J., Ferguson, A. M. N., Huxor, A. P., Tanvir, N. R., Ibata, R. A., \& Lewis, G. F. 2008, ApJL, 676, 17
\bibitem[Jeffery et al.(2011)]{jef11} Jeffery, E. J. et al. 2011, AJ, 141, 171
\bibitem[Jerjen et al.(1998)]{jfb98} Jerjen, H., Freeman, K. C., \& Binggeli, B. 1998, AJ, 116, 2873
\bibitem[Kallivayalil et al.(2006)]{kva06} Kallivayalil, N., van der Marel, R. P., Alcock, C., Axelrod, T., Cook, K. H., Drake, A. J., \& Geha, M. 2006, ApJ, 6
38, 772
\bibitem[Karachentsev et al.(2003)]{kar03} Karachentsev, D., et al. 2003, A\&A, 404, 93 
\bibitem[Karachentsev et al.(2004)]{kar04} Karachentsev, I. D., Karachentseva, V. E., Huchtmeier, W. K., \& Makarov, D. I. 2004, AJ, 127, 2031
\bibitem[Karachentsev et al.(2013)]{kar13} Karachentsev, I. D., Makarov, D. I., \& Kaisina, E. I. 2013, AJ, 145, 101
\bibitem[Kirby et al.(2008)]{ksg08} Kirby, E. N., Simon, J. D., Geha, M., Guhathakurta, P., \& Frebel, A. 2008, ApJL, 685, 43
\bibitem[Kobulnicky \& Koo (2000)]{kk00} Kobulnicky, H. A., \& Koo, D. C. 2000, ApJ, 545, 712
\bibitem[Kobulnicky et al.(2003)]{kob03} Kobulnicky, H. A. et al. 2003, ApJ, 599, 1006
\bibitem[Kobulnicky \& Kewley (2004)]{kk04} Kobulnicky, H. A., \& Kewley, L. J. 2004, ApJ, 617, 240
%\bibitem[Larmareille et al.(2004)]{lar04} Larmareille et al. 2004
\bibitem[Larson (1972)]{lar72} Larson, R. B. 1972, Nature, 236, 21
\bibitem[Layden (1998)]{lay98} Layden, A. C. 1998, AJ, 115, 193 
\bibitem[Lee \& Carney (1999)]{lc99} Lee, J.-W., \& Carney, B. W. 1999, AJ, 118, 1373
\bibitem[Lequex et al.(1979)]{lpr79} Lequeux, J., Peimbert, M., Rayo, J. F., Serrano, A., \& Torres-Peimbert, S. 1979, A\&A, 80, 155
\bibitem[Lianou et al.(2013)]{lgd13} Lianou, S., Grebel, E. K, Da Costa, G. S., Rejkuba, M., Jerjen, H., \& Koch, A. 2013, A\&A, 550, A7 (L13)
\bibitem[Madore \& Freedman (1995)]{mf95} Madore, B. F., \& Freedman, W. L. 1995, AJ, 109, 1645
\bibitem[Maier et al.(2004)]{mmh04} Maier, C., Meisenheimer, K., \& Hippelein, H. 2004, A\&A, 418, 475
\bibitem[Mackey \& Gilmore (2003)]{mg03} Mackey, A. D., \& Gilmore, G. F. 2003, MNRAS, 343, 747
\bibitem[Martin et al.(2009)]{mar09} Martin, N. F. et al. 2009, ApJ, 705, 758
\bibitem[Mateo (1998)]{mat98} Mateo, M. L. 1998, ARA\&A, 36, 435
\bibitem[Matteucci (2012)]{mat12} Matteucci, F. 2012, Chemical Evolution of Galaxies (Heidelberg: Springer)
\bibitem[McConnachie et al.(2008)]{mcc08} McConnachie, A. W. et al. 2008, ApJ, 688, 1009
\bibitem[Norris et al.(2010)]{nor10} Norris, J. E. et al. 2010, ApJ, 723, 1632
\bibitem[Oemler (1974)]{oem74} Oemler, A. J. 1974, ApJ, 194, 10
\bibitem[Pagel \& Patchett (1975)]{pp75} Pagel, B. E. J., \& Patchett, B. E. 1975, MNRAS, 172, 13
\bibitem[Pagel (1997)]{pag97} Pagel, B. E. J. 1997, Nucleosynthesis and Chemical Evolution of Galaxies (Cambridge, UK: Cambridge University Press)
, ISBN 0521550610
%\bibitem[]{} Pettini et al. 2001
\bibitem[Postman \& Geller (1984)]{pg84} Postman, M. \& Geller, M. J. 1984, ApJ, 281, 95
\bibitem[Pritzl et al.(2005)]{paj05} Pritzl, B. J., Armandroff, T. E., Jacoby, G. H., \& Da Costa, G. S. 2005, 129, 2232
\bibitem[Rejkuba et al.(2011)]{rhg11} Rejkuba, M., Harris, W. E., Greggio, L., \& Harris, G. L. H. 2011, A\&A, 526, 123
\bibitem[Richardson et al.(2011)]{ric11} Richardson, J. C. et al. 2011, ApJ, 732, 76
\bibitem[Sarajedini \& Jablonka (2005)]{sj05} Sarajedini, A., \& Jablonka, P. 2005, AJ, 130, 1627
\bibitem[Sarajedini et al.(2006)]{sbg06} Sarajedini, A., Barker, M. K., Geisler, D., Harding, P., \& Schommer, R. 2006, AJ, 132, 1361
\bibitem[Sarajedini et al.(2012)]{sym12} Sarajedini, A., Yang, S-C., Monachesi, A., Lauer, T. R., \& Trager, S. C. 2012, MNRAS, 425, 1459
\bibitem[Salaris et al.(1993)]{scs93} Salaris, M., Chieffi, A., \& Straniero, O. 1993, ApJ, 414, 580
\bibitem[Salzer et al.(2005)]{slm05} Salzer, J. J., Lee, J. C., Melbourne, J., Hinz, J. L., Alonso-Herrero, A., \& Jangren, A. 2005, ApJ, 624, 661
\bibitem[Savaglio et al.(2005)]{sav05} Savaglio, S. et al. 2005, ApJ, 635, 260
\bibitem[Schmidt (1963)]{sch63} Schmidt, M. 1963, ApJ, 137, 758
\bibitem[Schneider et al.(2002)]{sfn02} Schneider, R., Ferrara, A., Natarajan, P., \& Omukai, K. 2002, ApJ, 571, 30
\bibitem[Searle \& Sargent (1972)]{ss72} Searle, L., \& Sargent, W. L. W. 1972, ApJ, 173, 25
\bibitem[Schegel, Finkbeiner, \& Davis (1998)]{sfd98} Schlegel, D. J., Finkbeiner, D. P., \& Davis, M. 1998, ApJ, 500, 525
%\bibitem[]{} Shapley et al. 2004
\bibitem[Shapley et al.(2005)]{sha05} Shapley, A. E. et al. 2005, ApJ, 626, 698
\bibitem[Sirianni et al.(2005)]{sjb05} Sirianni, M., Jee, M. J., Bentez, N., \& et al. 2005, PASP, 117, 1049
\bibitem[Skillman et al.(1989)]{skh89} Skillman, E. D., Kennicutt, R. C., \& Hodge, P. W. 1989, ApJ, 347, 875
\bibitem[Slater et al.(2011)]{sbm11} Slater, C. T., Bell, E. F., \& Martin, N. F. 2011, ApJL, 742, 14
\bibitem[Tamman et al.(2003)]{tsr03} Tamman, G. A., Sandage, A., \& Reindl, B. 2003, A\&A, 404, 423
\bibitem[Timmes et al.(1995)]{tww95} Timmes, F. X., Woosley, S. E., \& Weaver, Thomas A. 1995, ApJS, 98, 617
\bibitem[Tolstoy et al.(2009)]{tht09} Tolstoy, E., Hill, V., \& Tosi, M. 2009, ARA\&A, 47, 371
\bibitem[Truran \& Cameron (1971)]{tc71} Truran, J. W., \& Cameron, A. G. W. 1971, Ap\&SS, 14, 179
\bibitem[van den Bergh (1962)]{van62} van den Bergh, S. 1962, AJ, 67, 486
\bibitem[van der Wel et al.(2010)]{van10} van der Wel, A., Bell, E. F., Holden, B. P., Skibba, R. A., Rix, H. 2010, ApJ, 714, 1779
\bibitem[Walsh et al.(2007)]{wjw07} Walsh, S. M., Jerjen, H., \& Willman, B. 2007, ApJL, 662, 83
\bibitem[Weisz et al.(2011)]{wei11} Weisz, D. R. et al. 2011, ApJ, 739, 5 (W11)
\bibitem[Wetzel et al.(2012)]{wtc12} Wetzel, A. R., Tinker, J. L., \& Conroy, C. 2012, MNRAS, 424, 232
\bibitem[Wheeler et al.(1989)]{wst89} Wheeler, J. C., Sneden, C., \& Truran, J. W., Jr. 1989, ARA\&A, 27, 279
\bibitem[Willman et al.(2005a)]{wil05a} Willman, B. et al. 2005a, AJ, 129, 2692
\bibitem[Willman et al.(2005b)]{wil05b} Willman, B. et al. 2005b, ApJL, 626, 85
\bibitem[Willman et al.(2011)]{wil11} Willman, B. et al. 2011, AJ, 142, 128
\bibitem[White \& Rees (1987)]{wr78} White, S. D. M., \& Rees, M. J. 1978, MNRAS, 183, 341	
\bibitem[White \& Frenk (1991)]{wf91} White, S. D. M., Frenk, C. S. 1991, ApJ, 379, 52
\bibitem[Wyse \& Gilmore (1993)]{wg93} Wyse, R. F. G., \& Gilmore, G. 1993, in ASP Conf Ser. 48, The globular clusters-galaxy connection, ed. G. H. Smith, \& J. P. Brodie (Santa Cruz, San Francisco, CA: ASP), 727
\bibitem[Yang et al.(2010)]{ysh10} Yang, S.-C., Sarajedini, A., Holtzman, J. A., \& Garnett, D. R. 2010, ApJ, 724, 799 (Y10)
\bibitem[Yang \& Sarajedini (2012)]{ys12} Yang, S.-C., \& Sarajedini, A. 2012, MNRAS, 419, 1362 (YS12)
\bibitem[Zaritsky et al.(1994)]{zkh94} Zaritsky, D., Kennicutt, R. C., Jr., \& Huchra, J. P. 1994, 1994, ApJ, 420, 87
\bibitem[Zinn \& West (1984)]{zw84} Zinn, R. \& West, M. 1984, ApJS, 55, 45
\bibitem[Zorotovic et al.(2010)]{zor10} Zorotovic, M. et al. 2010, AJ, 139, 357
\bibitem[Zucker et al.(2004)]{zuc04} Zucker, D. B. et al. 2004, ApJL, 612, 121
\bibitem[Zucker et al.(2006a)]{zuc06a} Zucker, D. B. et al. 2006a, ApJL, 643, 103
\bibitem[Zucker et al.(2006b)]{zuc06b} Zucker, D. B. et al. 2006b, ApJL, 650, 41
\bibitem[Zucker et al.(2007)]{zuc07} Zucker, D. B. et al. 2007, ApJL, 659, 21
\end{thebibliography}
\end{document}